\documentclass[10pt]{article}
\usepackage{graphicx}      
\usepackage[numbers]{natbib}        
\usepackage{graphicx}
\usepackage{algorithmic,algorithm}
\usepackage{amssymb,amsmath,amsthm}
\usepackage{mathrsfs}
\usepackage{xspace}
\usepackage{hyperref}
\usepackage{booktabs}
\usepackage{multicol}
\usepackage{bbm}

\usepackage[textwidth=4cm, textsize=footnotesize]{todonotes}
\usepackage{color}

\usepackage{lindsten,abbreviations}
\renewcommand\nonnegatives{\reals_+} 
\newcommand\pdf{PDF\@\xspace}

\newcommand\pgas{PGAS\@\xspace}
\newcommand\pgbs{PGBS\@\xspace}

\newcommand\TV{\operatorname{TV}}

\newcommand\ABCKernel[1]{\kappa_{#1}}

\newcommand\gibbsnum[1]{\emph{(#1)}}

\newcommand\xx{\mathbf{x}}
\renewcommand\aa{\mathbf{a}}

\newcommand\XX{\mathbf{X}}
\renewcommand\AA{\mathbf{A}}

\newcommand\tDist[2]{\gamma_{#2}}
\newcommand\ntDist[2]{\bar\gamma_{#2}}

\renewcommand\r[2]{r_{#2}}
\newcommand\wf[2]{\omega_{#2}}

\newcommand\ergoR[1][\Np]{R_{#1}}
\newcommand\ergorho[1][\Np]{\rho_{#1}}

\newcommand\ergokappa[1][\Np]{\kappa}

\usepackage{color}

\newtheoremstyle{ex}
  { }
  { }
  { }
  { }
  {\bfseries}
  { }
  { }
  {\thmname{#1}\thmnumber{ #2}:\thmnote{ #3}}
\theoremstyle{ex}

\theoremstyle{remark}
\newtheorem{remark}{Remark}

\theoremstyle{plain}

\newtheorem{theorem}{Theorem}

\newcounter{asmp}
\def\asmpfont{\upshape}

\graphicspath{{./}{./figs/}}

\DeclareFontFamily{U}{mathx}{\hyphenchar\font45}
\DeclareFontShape{U}{mathx}{m}{n}{
      <5> <6> <7> <8> <9> <10>
      <10.95> <12> <14.4> <17.28> <20.74> <24.88>
      mathx10
      }{}
\DeclareSymbolFont{mathx}{U}{mathx}{m}{n}
\DeclareFontSubstitution{U}{mathx}{m}{n}
\DeclareMathAccent{\widecheck}{0}{mathx}{"71}
\DeclareMathAccent{\wideparen}{0}{mathx}{"75}

\usepackage{tikz}
\usepackage{pgfplots}
 \usetikzlibrary{plotmarks,fit}
 \pgfplotsset{compat=newest}
 \pgfplotsset{plot coordinates/math parser=false}
 \usepgfplotslibrary{external}

\renewcommand\mid{\,\vert\,}
\newcommand\ma{\textbf{(M1)}\xspace}
\newcommand\mb{\textbf{(M2)}\xspace}

\newcommand\f{f}
\newcommand\g{g}

\newcommand\X{X} 
\newcommand\Y{Y} 
\newcommand\B{B} 
\newcommand\FX{\widetilde X} 
\newcommand\FB{\widetilde B} 
\newcommand\FY{\widetilde Y} 
\newcommand\UX{\Xi} 
\newcommand\RX{\widetilde \Xi} 
\newcommand{\ppw}[2]{\nu_{#1}^{#2}}
\newcommand\pd[1]{\phi_{#1}}
\newcommand\pdd[1]{\psi_{#1}} 
\newcommand\drv{v} 
\newcommand\extpdf{\pi^\Np_\T}
\newcommand\FXX{\widetilde \XX}
\newcommand\FAA{\widetilde \AA}
\newcommand\ct{\tau}
\newcommand\cX{Z} 
\newcommand\iX{\bar x} 
\newcommand\normind{j}
\newcommand\eqdef{:=}
\newcommand\eqdefi{=:}
\newcommand\rg[1]{\text{range}(#1)}

\title{Particle ancestor sampling for near-degenerate 
or intractable state transition models\thanks{Supported by the projects \emph{Learning of complex dynamical systems} (Contract number:
   637-2014-466) and \emph{Probabilistic modeling of dynamical systems} (Contract number: 621-2013-
   5524), both funded by the Swedish Research Council, and the project \emph{Bayesian Tracking and Reasoning over Time} (Reference: EP/K020153/1), funded by the EPSRC.}
}

\author{Fredrik Lindsten, Pete Bunch, Sumeetpal S. Singh,\\ and Thomas B. Sch{\"o}n}
\date{23 May 2015}

\begin{document}

\maketitle

\begin{abstract}
We consider Bayesian inference in sequential latent variable models in general, and in nonlinear
state space models in particular (\ie, state smoothing). We work with sequential Monte Carlo (SMC)
algorithms, which provide a powerful inference framework for addressing this problem.
However, for certain challenging and common model classes the state-of-the-art algorithms still struggle.
The work is motivated in particular by two such 
model classes: \emph{(i)} models where the state transition kernel is (nearly) degenerate, \ie
(nearly) concentrated on a low-dimensional manifold, and \emph{(ii)} models
where point-wise evaluation of the state transition density is intractable.
Both types of models arise in many applications of interest, including tracking, epidemiology,
and econometrics. The difficulties with these types of models is that they essentially
rule out forward-backward-based methods, which are known to be of great practical importance,
not least to construct computationally efficient
particle Markov chain Monte Carlo (PMCMC) algorithms.
To alleviate this, we propose a ``particle rejuvenation'' technique
to enable the use of the forward-backward strategy for (nearly) degenerate models and,
by extension, for intractable models.
We derive the proposed method 
specifically within the context of PMCMC, 
but we emphasise that it is applicable to any forward-backward-based Monte Carlo method.
\end{abstract}

\section{Problem formulation}
State space models (\ssm{s}) are widely used for modelling time series and
dynamical systems. A general, discrete-time \ssm can be written as
\begin{subequations}
  \label{eq:ssm}
  \begin{align}
    x_t\mid x_{t-1} &\sim \f(x_t \mid x_{t-1}), \\
    y_t\mid x_t &\sim \g(y_t \mid x_t),
  \end{align}
\end{subequations}
where $x_t \in \setX$ is the latent state, $y_t \in \setY$ is the observation (both at time~$t$) and
$\f(\cdot)$ and $\g(\cdot)$ are probability density functions (\pdf{s}) encoding the state transition and
the observation likelihood, respectively. 
The initial state is distributed according to $x_1 \sim \mu(x_1)$.

Statistical inference in \ssm{s} typically involves computation
of the \emph{smoothing distribution}, that is, the posterior distribution of a sequence
of state variables $\X_{\T} \eqdef \prange{x_1}{x_\T} \in \setX^\T$ conditionally on a
sequence of observations $\Y_{\T} \eqdef \prange{y_1}{y_\T} \in \setY^\T$.
The smoothing distribution plays
a key role both for offline (batch) state inference and for system identification via
data augmentation methods, such as expectation maximisation \cite{DempsterLR:1977} and Gibbs
sampling \cite{TannerW:1987}.

The motivation for the present work comes from two particularly challenging classes of \ssm{s}:

\vspace{1ex}\noindent\ma
If the state transition kernel $\f(\cdot)$ of the system puts all probability mass on some low-dimensional manifold,
we say that the transition is \emph{degenerate} 
(for simplicity we use ``probability density notation'' even in
the degenerate case). Degenerate transition kernels arise, \eg, if the dynamical evolution is modeled using
additive process noise with a rank-deficient covariance matrix. Models of this type are common in certain
application areas, \eg, navigation and tracking; see \cite{GustafssonGBFJKN:2002} for several examples.
Likewise, if $\f(\cdot)$ is concentrated around a low-dimensional manifold (\ie, the transition is highly informative,
or the process noise is small) we say that the transition is \emph{nearly degenerate}. 

\vspace{1ex}\noindent\mb
If the state transition density function $\f(\cdot)$ is not available on closed form,
the transition is said to be \emph{intractable}. The typical scenario is that
$\f(\cdot)$ is a regular (non-degenerate) \pdf which it is possible to simulate from,
but which nevertheless is intractable.
At first, this scenario might seem contrived, but it is in fact quite common in practice.
In particular, whenever the dynamical function is defined implicitly by some computer program or black-box simulator,
or as a ``complicated'' nonlinear transformation of a known noise input, it is typically
not possible to explicitly write down the corresponding transition \pdf; see, \eg, \cite{MurrayJP:2013,AndrieuDH:2010} for examples.

\vspace{1ex}
The main difficulty in performing inference for model classes \ma and \mb
lies in that the so-called \emph{backward kernel} of the model (see, \eg, \cite{LindstenS:2013}),
\begin{align}
  \label{eq:bkg:bwd-kernel}
  p(x_t \mid x_{t+1}, y_{1:t}) \propto \f(x_{t+1} \mid x_t) p(x_t \mid y_{1:t}),
\end{align}
will also be (nearly) degenerate or intractable, respectively.
This is problematic since many state-of-the-art methods rely on the backward kernel for inference;
see, \eg, \cite{GodsillDW:2004,LindstenJS:2014,Whiteley:2010}.
In particular, the backward kernel is used to implement the well-known
forward-backward smoothing strategy.
We will come back to this in the subsequent sections when we discuss the details of these
inference methods and how we propose an extension to the methodology geared toward this issue.


The main contribution of this paper is constituted by a construction allowing us to replace the
simulation from the (problematic) backward kernel, with simulation from a joint density over a
subset of the future state variables. 
Importantly, this joint density will typically have more favourable properties than the backward
kernel itself, whenever the latter is (nearly) degenerate.  Simulating from the joint density
results in a \emph{rejuvenation} of the state values, 
which intuitively can be understood as a bridging between past and future state variables.
Furthermore, to extend the scope of this technique we propose a nearly degenerate approximation of an
intractable transition model.  Using this approximation we can thus use the proposed particle
rejuvenation strategy to perform inference also in models with intractable transitions.


\section{Background, methodology, and related work}
\subsection{Computational inference}
The strong assumptions of linearity and Gaussianity that were originally invoked
for state space inference have been significantly weakened by the development of computational
statistical methods. Among these, Markov chain Monte Carlo (\mcmc) and sequential Monte Carlo 
(\smc, a.k.a.\@\xspace particle filter) methods play prominent roles.
\mcmc methods (see \eg, \cite{RobertC:2004,AndrieuFDJ:2003}) are based on simulating a Markov chain
which admits the target distribution of interest---here, the joint smoothing distribution---as its
unique stationary distribution.
\smc methods, on the other hand (see, \eg, \cite{DoucetJ:2011,DelMoralDJ:2006,DelMoral:2004}),
use a combination of sequential importance sampling  \cite{HandschinM:1969} and resampling \cite{Rubin:1987}
to approximate a sequence of probability distributions defined
on a sequence of measurable spaces of increasing dimension. This is accomplished by
approximating each target distribution by an empirical point-mass distribution
based on a collection of random samples, referred to as \emph{particles},
with corresponding non-negative \emph{importance weights}.
For instance, the target distributions of an \smc sampler can comprise the smoothing distributions
for an \ssm for $t=\range{1}{\T}$ and, indeed, \smc methods have emerged as a key tool for approximating the flow of smoothing
distributions for general \ssm{s}.

In addition to these methods, we have in recent years seen much interest in the combination of \smc and \mcmc in
so-called particle \mcmc (\pmcmc) methods \cite{AndrieuDH:2010}. These methods are based on an
auxiliary variable construction which opens up the use of \smc (or variants thereof)
to construct efficient, high-dimensional \mcmc transition kernels.
The introduction of \pmcmc has spurred intensive research in the community spanning methodological
\cite{CarterMK:2014,LindstenJS:2014,WhiteleyAD:2010,OlssonR:2010},
theoretical \cite{LindstenDM:2015,AndrieuLV:2013,DelMoralKP:2014,ChopinS:2014,DoucetPDK:2014,AndrieuV:2012},
and applied \cite{VrugtBDS:2013,PittSGK:2012,GolightlyW:2011,RasmussenRK:2011} work.

In this paper we consider in particular the \emph{particle Gibbs} (\pg) algorithm, introduced by \citet{AndrieuDH:2010}.
The \pg algorithm relies on running a modified particle filter, in which one particle trajectory is
set deterministically according to an \emph{a priori} specified reference trajectory (see Section~\ref{sec:pgas}).
After a complete run of the \smc algorithm, a new trajectory
is obtained by selecting one of the particle trajectories with probabilities given by their
importance weights. This results in a Markov kernel on the \emph{space of trajectories} $\setX^\T$.
Interestingly, the conditioning on a reference trajectory ensures that the limiting distribution of this Markov kernel
is exactly the target distribution of the sampler (\eg, the joint smoothing distribution)
for any number of particles $\Np\geq 2$ used in the underlying \smc algorithm.
The \pg algorithm can be interpreted as a Gibbs sampler for the extended model where the random variables generated by
the \smc sampler are treated as auxiliary variables.

Like any Gibbs sampler, \pg has the advantage over Metropolis-Hastings of not requiring an accept/reject stage.
However, the resulting chain is still liable to mix (very) slowly if the particle filter suffers from path-space
degeneracy \cite{ChopinS:2014,LindstenS:2013}. Unfortunately, path degeneracy is inevitable
for high-dimensional (large~$\T$) problems, which significantly reduces the applicability of \pg.
This problem has been addressed in a generic setting by adding additional sampling steps to the \pg sampler,
either during the filtering stage, known as particle Gibbs with ancestor sampling (\pgas) \cite{LindstenJS:2014}, or in an
additional backward sweep, known as particle Gibbs with backward simulation (\pgbs) \cite{WhiteleyAD:2010,Whiteley:2010}.
The improvement arises from sampling new values for individual particle ancestor indexes, and thus allowing
the reference trajectory to be updated gradually which can mitigate the effects of path degeneracy.
It has been found that this can \emph{vastly} 
improve mixing, making the method much more robust to a
small number of particles as well as growth in the size of the data.

\subsection{Tackling (nearly) degenerate or intractable transitions}
A problem with \pgas and \pgbs, however, is that they rely on a particle approximation of
the \emph{backward kernel} for updating the ancestry of the particles. For an \ssm the backward kernel is given by \eqref{eq:bkg:bwd-kernel}, 
which implies that if $\f(\cdot)$ is (nearly) degenerate or intractable, then so is the backward kernel.
As an effect, the probability of sampling any change in the particle ancestry
is low. This largely removes the effect of the extra sampling steps introduced in order to reduce path
degeneracy. In fact, if the transition is truly degenerate, then the probability of updating the ancestry will be exactly zero.
Intuitively, the problem is that the only state history consistent with a particular ``future state'' is that from which
the future state was originally generated.

To mitigate this effect we propose to use a procedure which we call \emph{particle rejuvenation}. A specific instance of this technique has previously been used together with forward/backward particle smoothers
\cite{BunchG:2013}---when sampling an ancestor index they simultaneously sample a new value for the associated state.
In the preliminary work \cite{BunchLS:2015} we investigated the effect of this approach on the \pgbs sampler.
This opened up for steering the potential state histories towards the fixed future, consequently increasing the probability of changing the ancestry and thus improving the mixing of the Markov chain.
Independently, \citet{CarterMK:2014} have derived essentially the same method. However, they
introduce a different extended target distribution in order to justify this addition, which
necessitates changes to the underlying \smc sampler, whereas our developments allows us to use the
standard \pmcmc construction.

In the present work we extend the method from \cite{BunchLS:2015,CarterMK:2014} by considering a more generic setting,
in which we propose to rejuvenate any ``subset'' of the reference trajectory along with the particle ancestry
(Section~\ref{sec:pgras}).
This provides additional flexibility over previous approaches. As we illustrate
in Section~\ref{sec:degen}, this
increased flexibility is necessary in order to address the challenges associated with several models
of interest containing \mbox{(nearly)} degenerate transitions (type \ma).
Furthermore, in Section~\ref{sec:abc} we show how an intractable transition can be approximated with a nearly degenerate one.
Employing the particle rejuvenation strategy to this approximation 
results in inference methods applicable to models of type \mb where $\f(\cdot)$, and thus the backward kernel~\eqref{eq:bkg:bwd-kernel},
is intractable.
Here, again, the increased flexibility provided by the generalised particle rejuvenation strategy
is key. 
%
We work specifically with the \pgas algorithm, but all the proposed modifications are also applicable
to \pgbs and, in fact, to any other backward-simulation-based method (see~\cite{LindstenS:2013}).

\section{Particle Gibbs with Ancestor Sampling}\label{sec:pgas}
%
%
%
While \ssm{s} comprise one of the main motivating factors for the
development of \smc and \pmcmc (and we shall return to these models in the sequel), these methods
can be applied more generally, see, \eg, \cite{AndrieuDH:2010,LindstenS:2013}. For this reason we will
carry out the derivation in a more general setting.

We start by reviewing the \pgas algorithm by \citet{LindstenJS:2014}.
%
Consider the problem of sampling from a possibly high-dimensional probability distribution.
As before, we write $\X_{\T} \eqdef \prange{x_1}{x_\T} \in \setX^\T$ for the random variables of the model.
%
Let $\tDist{\theta}{t}(\X_{t})$ for $t = \range{1}{\T}$ be a sequence of unnormalised densities on $\setX^t$,
which we assume can be evaluated point-wise. Let $\ntDist{\theta}{t}(\X_{t})$ be the corresponding
normalised \pdf{s}. The central object of interest is assumed to be the final \pdf in the sequence:
$\ntDist{\theta}{\T}(\X_{\T})$.
For an \ssm we would typically have $\ntDist{\theta}{t}(\X_{t}) = p(\X_{t} \mid \Y_{t})$
and $\tDist{\theta}{t}(\X_{t}) = p(\X_{t}, \Y_{t})$.

The \pgas algorithm \cite{LindstenJS:2014} is a procedure for constructing a Markov kernel on $\setX^\T$ which
admits $\ntDist{\theta}{\T}(\X_{\T})$ as its unique stationary distribution. Consequently, by simulating
a Markov chain with transition probability given by the, so-called, \pgas kernel, we will (after an initial transient phase)
obtain samples distributed according to $\ntDist{\theta}{\T}(\X_{\T})$.
The \pgas kernel can thus readily be used in \mcmc and related Monte Carlo methods based on Markov kernels.
As mentioned above, \pgas is an instance of \pmcmc. Specifically, the construction
of the \pgas kernel relies on (a variant of) an \smc sampler, targeting the sequence of intermediate
distributions $\tDist{\theta}{t}(\X_{t})$ for $t = \range{1}{\T}$.

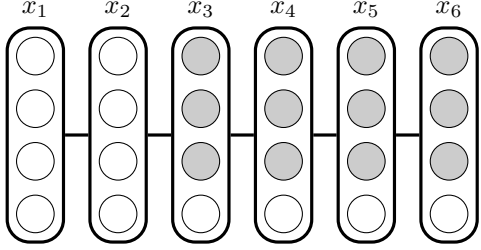
\begin{figure}[ptb]
  \centering
  \hspace*{-1.5em}\tikzstyle{edge} = [-]
\tikzstyle{edge2} = [-,very thick,>=latex]
\tikzstyle{arrw} = [very thick,shorten <=2pt,shorten >=2pt]
\tikzstyle{var} = [draw,circle,inner sep=0,minimum width=0.5cm]
\tikzstyle{obs} = [draw,circle,inner sep=0,minimum width=0.5cm, fill=black!20]
  \begin{tikzpicture}[>=stealth,node distance=0.6cm]
    \begin{scope}
      \foreach \x in {0,1} {
        \foreach \y in {0,1,2,3} {
          \node at (1.1*\x,0.7*\y) (x\x\y) [var] {};
        }
      }
      \foreach \x in {2,3,4,5} {
        \foreach \y in {0} {
          \node at (1.1*\x,0.7*\y) (x\x\y) [var] {};
        }
      }
      \foreach \x in {2,3,4,5} {
        \foreach \y in {1,2,3} {
          \node at (1.1*\x,0.7*\y) (x\x\y) [obs] {};
        }
      }
      \foreach \x in {0,1,2,3,4,5} {
        \pgfmathtruncatemacro\xend{\x+1}
        \node[draw,very thick,rectangle,rounded corners=3mm,minimum width=0.6cm,fit=(x\x0) (x\x3),label=above:{$x_{\xend}$}] (x\x){};
      }
      \foreach \x in {0,1,2,3,4} {
        \pgfmathtruncatemacro\xend{\x+1}
          \draw[edge2] (x\x) -> (x\xend);
      }
    \end{scope}
  \end{tikzpicture}
  \caption{Illustration of subset notation $\UX \subset \X_\T$. Here, $\setX = \reals^4$ and each component of
  the vector $x_t$ is illustrated by a circle. The gray disks illustrate a subset $\UX \subset \X_6$,
  consisting of $\UX = \{ x_{j,t} : j=1,\,2,\,3 \text{ and } t = 3,\,4,\,5,\,6\}.$}
  \label{fig:pgas:set-notation}
\end{figure}

For later reference we introduce some additional notation. First, we will make frequent use of the
notation already exemplified above, where capital letters are used for sequences of variables, \eg,
by writing $\X_t$ for the ``past'' history of the state sequence.  Similarly,
$\FX_{t+1} \eqdef \X_\T \setminus \X_{t} = \prange{x_{t+1}}{x_\T}$ denotes the ``future'' state
sequence. Here, we have used \emph{set notation} to refer to a collection of latent variables and
we will make frequent use also of this notation in the sequel. In particular, we will write
$\UX \subseteq \X_\T$ for an arbitrary subset of the latent random variables of the model,
which could refer to individual components of $x_t$, say, when $x_t$ is a vector (see
Section~\ref{sec:abc}). For clarity, we exemplify this usage of set notation in Figure~\ref{fig:pgas:set-notation}.

The \pgas algorithm is reminiscent of a standard \smc sampler, but with the important
difference that one particle at each iteration is specified \emph{a priori}.
These particles, denoted as $\prange{x_1^\prime}{x_\T^\prime} \eqdefi \X_\T^\prime$
and with corresponding particle indexes $\prange{b_1}{b_\T} \eqdefi \B_\T$,
serve as a reference for the sampler, as detailed below.



As in a standard \smc sampler, we approximate the sequence of target densities
$\ntDist{\theta}{t}(\X_t)$ for $t = \range{1}{\T}$ by collections of weighted particles.
Let $\{\X_{t-1}^i, w_{t-1}^i\}_{i=1}^\Np$ be a particle system approximating
$\ntDist{\theta}{t-1}(\X_{t-1})$ by the empirical distribution,
\begin{align}
 \ntDist{\theta}{t-1}^\Np(\X_{t-1}) \eqdef \sum_{i=1}^\Np \frac{ w_{t-1}^i}{ \sum_{\normind=1}^\Np w_{t-1}^\normind} \delta_{\X_{t-1}^i }(\X_{t-1}).
\end{align}
Here, $\delta_z(x)$ is a point mass located at $z$.
To propagate this particle system to time $t$, we introduce the auxiliary variables $\{a_t^i\}_{i=1}^\Np$,
referred to as \emph{ancestor indexes}, encoding the genealogy of the particle system.
More precisely, $x_t^i$ is generated by first sampling
its ancestor,
\begin{align}
  \label{eq:bkg:resampling}
  \Prb(a_t^i = j) &\propto w_{t-1}^j, &j = \range{1}{\Np}.
\end{align}
Then, $x_t^i$ is drawn from some proposal distribution,
\begin{align}
  \label{eq:bkg:propagation}
  x_t^i \sim \r{\theta}{t}(\cdot \mid \X_{t-1}^{a_t^i}).
\end{align}
When we write $\X_{t}^i$ we refer to the ancestral path of particle $x_t^i$.
That is, the particle trajectory is defined recursively as
\begin{align}
  \label{eq:bkg:trajectory-extension}
  \X_{t}^i  \eqdef (\X_{t-1}^{a_t^i}, x_t^i).
\end{align}
In this formulation the resampling step is implicit and corresponds to sampling the ancestor indexes.
Finally, the particle is assigned a new importance weight: $w_t^i = \wf{\theta}{t}(\X_{t}^i)$ where the weight
function is given by
\begin{align}
  \label{eq:smc:weightfunction}
  \wf{\theta}{t}(\X_{t}) = \frac{\tDist{\theta}{t}(\X_{t})}{ \tDist{\theta}{t-1}(\X_{t-1}) \r{\theta}{t}(x_t \mid \X_{t-1})}.
\end{align}
In a standard \smc sampler, the procedure above is repeated for each $i = \range{1}{\Np}$, to generate $\Np$
particles at time $t$. However, as mentioned above, the \pgas procedure relies on keeping
one particle fixed at each iteration in order to obtain the correct limiting distribution of the resulting
\mcmc kernel. This is accomplished by simulating according to \eqref{eq:bkg:resampling} and \eqref{eq:bkg:propagation}
only for $i \in \crange{1}{\Np} \setminus b_t$. The final particle is set deterministically according to the
reference: $x_t^{b_t} = x_t^\prime$.

To be able to construct the particle trajectory $\X_t^{b_t}$ as in
\eqref{eq:bkg:trajectory-extension}, the reference particle has to be associated with an ancestor at time $t-1$.
This is done by \emph{ancestor sampling}; that is, we simulate randomly a value for the corresponding
ancestor index $a_t^{b_t}$. \citet{LindstenJS:2014} derive the ancestor sampling distribution,
and show that $a_t^{b_t}$ should be simulated according to,
\begin{align}
  \label{eq:pgas:original_as}
  \Prb(a_t^{b_t} = i) &\propto  w_{t-1}^i \frac{\tDist{\theta}{\T}( \X_{t-1}^i \cup \FX_t^\prime  )}{ \tDist{\theta}{t-1} (\X_{t-1}^i) }, &i = \range{1}{\Np}.
\end{align}
Note that the expression depends on the complete ``future'' reference path $\FX_t^\prime = \prange{x_t^\prime}{x_\T^\prime}$.
In the above, $\X_{t-1}^i \cup \FX_t^\prime$ refers to the complete path formed by concatenating the
two partial trajectories.

The expression above can be understood as an application of Bayes' theorem; the weight $w_{t-1}^i$ is the prior probability of
particle $\X_{t-1}^i$ and the density ratio corresponds to the likelihood of ``observing'' $\FX_t^\prime$ given $\X_{t-1}^i$.
The actual motivation for the ancestor sampling distribution \eqref{eq:pgas:original_as}, however,
relies on a collapsing argument~\cite{LindstenJS:2014}.
We will review this idea in the subsequent section where we propose a generalisation
of the technique to increase the flexibility of the \pgas algorithm.

Finally, after a complete pass of the modified \smc procedure outlined above for $t = \range{1}{\T}$,
a new trajectory $\X_\T^\star$ is sampled by selecting among the particle trajectories
with probability given by their corresponding importance weights. That is, we sample $k$
with $\Prb(k=i) \propto w_\T^i$, $i = \range{1}{\Np}$ and return $\X_\T^\star = \X_\T^k$.
We summarise the \pgas procedure in Algorithm~\ref{alg:pgas-original}.
Note that the algorithm stochastically simulates the trajectory $\X_T^\star$ conditionally on
the reference trajectory $\X_\T^\prime$, thus implicitly defining a Markov kernel on $\setX^\T$.

\begin{remark}
  As pointed out in \cite{ChopinS:2014,LindstenJS:2014}, the indexes $\B_\T$ are nuisance variables
  that are unnecessary from a practical point of view, \ie, it is possible to set $\B_T$ to any
  arbitrary convenient sequence, \eg $B_\T = \prange{\Np}{\Np}$. We use this convention in
  Algorithms~\ref{alg:pgas-original}~and~\ref{alg:pgras}.
  However, explicit reference to the index variables will simplify the derivation of the algorithm.
\end{remark}

\begin{algorithm}
  \caption{\pgas Markov kernel \cite{LindstenJS:2014}}
  \label{alg:pgas-original}
  \begin{algorithmic}[1]
    \REQUIRE Reference trajectory $\X_\T^\prime \in \setX^\T$.
    \STATE Set $x_1^{\Np} = x_1^\prime$.
    \STATE Draw $x_1^i \sim \r{\theta}{1}(\cdot)$ for $i = \range{1}{\Np-1}$.
    \STATE Set $w_1^i = \tDist{\theta}{1}(x_1^i)/\r{\theta}{1}(x_1^i)$ for ${i = \range{1}{\Np}}$.
    \FOR{$t = 2$ \TO $\T$}
    \STATE Simulate $( a_t^i, x_t^i)$ as in (\ref{eq:bkg:resampling}, \ref{eq:bkg:propagation}) for ${i =\range{1}{\Np-1}}$.
    \STATE Simulate $a_t^\Np$ according to \eqref{eq:pgas:original_as}.
    \STATE Set $x_t^{\Np} = x_{t}^\prime$.
    \STATE Set $\X_{t}^i = ( \X_{t-1}^{a_t^i}, x_t^i )$ for $i = \range{1}{\Np}$.
    \STATE Set $w_t^i = \wf{\theta}{t}(\X_{t}^{i})$ for $i = \range{1}{\Np}$.
    \ENDFOR
    \STATE Draw $k$ with $\Prb(k = i) \propto w_\T^i$.
    \RETURN $\X_{\T}^\star = \X_{\T}^k$.
  \end{algorithmic}
\end{algorithm}

\begin{remark}
  \label{rem:as-important}
  PGAS is a variation of the \pg sampler by \citet{AndrieuDH:2010}.  Algorithmically, the
  only difference between the methods lies in the ancestor sampling
  step~\eqref{eq:pgas:original_as}, where, in the original \pg sampler we would simply
  set $a_t^\Np = \Np$ (deterministically). While this is a small modification, it can have a very
  large impact on the convergence speed of the method, as discussed and illustrated in
  \cite{LindstenJS:2014}. Informally, the reason for this is that the \emph{path degeneracy} of SMC
  samplers will cause the \pg sampler to degenerate toward the reference trajectory. That
  is, with high probability $x_s^\star = x_s^\prime$ for any $s \ll \T$, effectively causing the
  sampler to be stuck at its initial value in certain parts of the state space $\setX^\T$. Ancestor
  sampling mitigates this issue by assigning a new ancestry of the reference trajectory
  via~\eqref{eq:pgas:original_as}.  Path degeneracy will still occur, but the sampler tends to
  degenerate toward something different than the reference trajectory ($x_s^\star \neq x_s^\prime$
  with non-negligible probability) enabling much more efficient exploration of the state space.
\end{remark}

\section{Particle rejuvenation for the PGAS kernel} \label{sec:pgras}
We now turn to the new procedure: a \emph{particle rejuvenation} strategy for PGAS.
The idea is to update the ancestor index $a_t^{b_t}$ jointly with some subset $\UX_t \subseteq \FX_t^\prime$
of the reference trajectory. The intuitive motivation is that by ``loosening up'' the reference
trajectory, we get more freedom of changing its ancestry.
Before we continue with the derivation of the particle rejuvenation strategy, however, we
shall consider a simple numerical example to motivate the development.

\subsection{An illustrative example} \label{sec:pgras:motivating} 
Suppose we wish to track the motion of an object in 3D space from noisy measurements of its
position. For the transition model $f_\theta(x_t \mid x_{t-1})$ we assume near constant velocity
motion \cite{Li2003}, and the observation model is noisy measurements of bearing, elevation and
range. Here we have also introduced an unknown system parameter $\theta$
corresponding to 
the scale factor on the transition covariance matrix,
which characterises the target manoeuvrability.
Based on a batch of $\T = 100$ observations $\Y_\T$, we wish to learn the unknown parameter
$\theta$ by Gibbs sampling, iteratively simulating from $p(\theta \mid \X_\T, \Y_\T)$
and $p(\X_\T \mid \theta, \Y_\T)$. (See \cite{BunchLS:2015} for more details on the
model and experimental setup.)

This model can be problematic for \pgas if the scale parameter $\theta$ is small, since this implies
a highly informative, \ie,
nearly degenerate, transition model. To be more precise, for an \ssm, with the unnormalised target density being $\gamma_{\theta,t}(\X_t) = p(\X_t, \Y_t \mid \theta)$, it follows that the ancestor sampling distribution
in \eqref{eq:pgas:original_as} simplifies to
\begin{align}
  \label{eq:pgras:tracking_as}
  \Prb(a_t^{b_t} = i) &=  \frac{ w_{t-1}^i f_\theta(x_{t}^\prime \mid x_{t-1}^i) }{ \sum_{\normind=1}^\Np w_{t-1}^\normind f_\theta(x_{t}^\prime \mid x_{t-1}^\normind) }, &i = \range{1}{\Np}.
\end{align}
Now, if $\theta$---the process noise---is small, then $f_\theta(x_t \mid x_{t-1})$ is largely concentrated on a single point. Hence, the distribution in \eqref{eq:pgras:tracking_as} will also be
concentrated on one value and there is little freedom in changing the ancestry of $x_t^\prime$ at time $t-1$.
The result is that the effect of ancestor sampling is diminished.

The idea with particle rejuvenation is that simultaneously sampling a new state $x_t^\prime$, for instance, jointly with the ancestor index 
opens up for \emph{bridging} between the states
$x_{t-1}^i$
and $x_{t+1}^\prime$.
This leads to a substantially higher probability of updating the ancestor indexes during each iteration, and hence faster mixing.

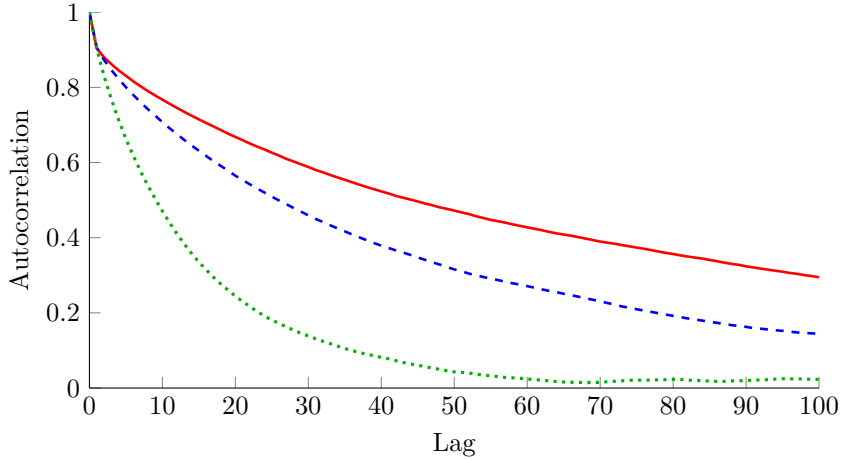
\begin{figure}
%
%
\begin{tikzpicture}

\begin{axis}[%
width=0.8\linewidth,
height=5cm,
scale only axis,
xmin=0,
xmax=100,
xlabel={Lag},
ymin=0,
ymax=1,
ylabel={Autocorrelation},
axis x line*=bottom,
axis y line*=left,
]
\addplot [
color=red,
line width=1pt,
solid
]
table[row sep=crcr]{
0 1\\
1 0.904022735448337\\
2 0.880033906595486\\
3 0.862092002864626\\
4 0.845224105659387\\
5 0.830899907238224\\
6 0.816205957449242\\
7 0.80307821173986\\
8 0.790318391490336\\
9 0.778743462976878\\
10 0.767299542809978\\
11 0.756128215498345\\
12 0.745409051529832\\
13 0.734788609466434\\
14 0.724757758492534\\
15 0.714975202099996\\
16 0.705407593020561\\
17 0.696141550314619\\
18 0.686550564623964\\
19 0.676883228286137\\
20 0.668187562302004\\
21 0.659559901812037\\
22 0.650491270642776\\
23 0.642308805013972\\
24 0.63469574172827\\
25 0.626243205096502\\
26 0.618426585549508\\
27 0.609731758107273\\
28 0.602619865545163\\
29 0.595300797408976\\
30 0.588174556166848\\
31 0.580288071494473\\
32 0.57400124031902\\
33 0.567018644109268\\
34 0.56043698091566\\
35 0.554089420346626\\
36 0.547637332850567\\
37 0.541136713811495\\
38 0.53490812411435\\
39 0.529042621609207\\
40 0.523238762996096\\
41 0.517667994478097\\
42 0.511335615567858\\
43 0.506346025294487\\
44 0.501504251525869\\
45 0.496427692169461\\
46 0.491009080224316\\
47 0.486391076656039\\
48 0.480849020370124\\
49 0.47669225509172\\
50 0.472254369904352\\
51 0.467195168768281\\
52 0.462876889647369\\
53 0.45715147947141\\
54 0.452303712436505\\
55 0.447442669600364\\
56 0.443766910375299\\
57 0.440165329444226\\
58 0.435460594597378\\
59 0.431443224026525\\
60 0.427632155778983\\
61 0.423870704087662\\
62 0.419986045738758\\
63 0.415387586850307\\
64 0.411552536284577\\
65 0.408386425965555\\
66 0.405313684791632\\
67 0.401538057850948\\
68 0.397639991514691\\
69 0.393779088082263\\
70 0.389634356801077\\
71 0.386745094711647\\
72 0.383980523442736\\
73 0.380320651035342\\
74 0.377052535110575\\
75 0.373835921330528\\
76 0.370514242128762\\
77 0.366929405591482\\
78 0.363198075511641\\
79 0.359581525301984\\
80 0.356567025916033\\
81 0.35297794581045\\
82 0.349898978172386\\
83 0.347047305562659\\
84 0.344447138252992\\
85 0.340837563885843\\
86 0.337561676983212\\
87 0.333647761233221\\
88 0.3303263988232\\
89 0.327746299751151\\
90 0.323843788981056\\
91 0.320780681068063\\
92 0.317831189852745\\
93 0.31495560189811\\
94 0.312135805578388\\
95 0.309441320388971\\
96 0.306149625110323\\
97 0.303831213593552\\
98 0.300479825992381\\
99 0.29784682184687\\
100 0.294866382032209\\
};

\addplot [
color=blue,
line width=1pt,
dashed
]
table[row sep=crcr]{
0 1\\
1 0.905792878882163\\
2 0.873074705031518\\
3 0.846378375602707\\
4 0.822798904506012\\
5 0.800533117947232\\
6 0.780097783317\\
7 0.760700082736565\\
8 0.742105401802764\\
9 0.725098224058954\\
10 0.707372385457198\\
11 0.691080312230355\\
12 0.675852768391729\\
13 0.660029917912106\\
14 0.645462891274177\\
15 0.631493004836248\\
16 0.61743579953908\\
17 0.603910646164206\\
18 0.591118680710257\\
19 0.578593744851273\\
20 0.564932379557952\\
21 0.553492207178705\\
22 0.541867881597295\\
23 0.530796991904181\\
24 0.519952617887766\\
25 0.509265016419075\\
26 0.498823399349585\\
27 0.488988747700827\\
28 0.478982875867495\\
29 0.469369133447967\\
30 0.459352229200461\\
31 0.450057545065143\\
32 0.441813635068685\\
33 0.433076168717971\\
34 0.42479710512961\\
35 0.416943181810551\\
36 0.408592447865454\\
37 0.400828340003762\\
38 0.393342697297914\\
39 0.385850500573174\\
40 0.378647078263704\\
41 0.372655537708904\\
42 0.366163902438759\\
43 0.360385156485168\\
44 0.354250072072272\\
45 0.347434864334979\\
46 0.341148274354618\\
47 0.334694332004562\\
48 0.328568983911832\\
49 0.321983390416355\\
50 0.315644905020758\\
51 0.310759445351894\\
52 0.304527877824596\\
53 0.299917868514913\\
54 0.295394439624545\\
55 0.291004588102203\\
56 0.287220575177909\\
57 0.282738808112174\\
58 0.278730992435166\\
59 0.275181896443015\\
60 0.271098875607067\\
61 0.266983664868412\\
62 0.262412256024592\\
63 0.258674101421817\\
64 0.255042908844846\\
65 0.250836748623899\\
66 0.247102627421719\\
67 0.243110883598519\\
68 0.238730427185936\\
69 0.234900093652543\\
70 0.230743554496934\\
71 0.226231442811001\\
72 0.222160915538413\\
73 0.217007429566034\\
74 0.213329004555655\\
75 0.209626364664693\\
76 0.205987805352776\\
77 0.202625089167047\\
78 0.198863673333585\\
79 0.195368680068564\\
80 0.192050360212801\\
81 0.188613755752867\\
82 0.185168633573012\\
83 0.182381207000657\\
84 0.179403635404978\\
85 0.176288668179836\\
86 0.173534515348557\\
87 0.170294102929758\\
88 0.167450623165975\\
89 0.165090967978907\\
90 0.162791146495416\\
91 0.15939939010056\\
92 0.157767371158756\\
93 0.15583915661608\\
94 0.153819347034653\\
95 0.152167391861015\\
96 0.149984439717101\\
97 0.147901243860999\\
98 0.146230848925432\\
99 0.145588548159335\\
100 0.143723515814402\\
};

\addplot [
color=black!30!green,
line width=1.2pt,
dotted
]
table[row sep=crcr]{
0 1\\
1 0.902878729046372\\
2 0.830603648348448\\
3 0.76879414797804\\
4 0.713557251849928\\
5 0.662952701151596\\
6 0.618371908320766\\
7 0.577055217571375\\
8 0.538951043102581\\
9 0.503888732691735\\
10 0.470853642004247\\
11 0.439617322209893\\
12 0.409963264479712\\
13 0.38302486369425\\
14 0.35841257132926\\
15 0.335237746456692\\
16 0.314732930942974\\
17 0.294738073411605\\
18 0.27688672156697\\
19 0.260626985420551\\
20 0.244575416757429\\
21 0.230200351265455\\
22 0.21685119467824\\
23 0.203490137654066\\
24 0.191599126945049\\
25 0.180657685626003\\
26 0.172021225681158\\
27 0.162436492691492\\
28 0.153996694773004\\
29 0.145577870479578\\
30 0.138391598264147\\
31 0.131218718469463\\
32 0.124512967599903\\
33 0.118211797551436\\
34 0.111487896992776\\
35 0.105175351783038\\
36 0.0997534567250185\\
37 0.094428116054682\\
38 0.0897737746962458\\
39 0.0861556200417014\\
40 0.0813767086259135\\
41 0.0779222903042463\\
42 0.0727900756197038\\
43 0.0679391177465129\\
44 0.0641458915755258\\
45 0.060820201933026\\
46 0.056860414746139\\
47 0.0532505042362372\\
48 0.0495999333705118\\
49 0.0461293420439188\\
50 0.0430272724985708\\
51 0.0416834392107168\\
52 0.0399802407780156\\
53 0.0369417187380415\\
54 0.0342245874135671\\
55 0.0326787847698156\\
56 0.0303522185998093\\
57 0.0281371546795945\\
58 0.0271507748117603\\
59 0.025916549442171\\
60 0.0242364191104742\\
61 0.0220973685157745\\
62 0.0210714835130998\\
63 0.0191103645733615\\
64 0.0171714470880262\\
65 0.0161152775472994\\
66 0.0153526655139717\\
67 0.0150504594051858\\
68 0.0143382574454141\\
69 0.0146135242262952\\
70 0.0150623191062795\\
71 0.0168726891087079\\
72 0.0180632355692352\\
73 0.0190747290505219\\
74 0.0205862162363312\\
75 0.0211684968209856\\
76 0.02072558904443\\
77 0.0214485944357785\\
78 0.021659502929925\\
79 0.022260053721592\\
80 0.0231897527637606\\
81 0.0226799305531637\\
82 0.0212216978213011\\
83 0.0200557936227753\\
84 0.0198397251547976\\
85 0.0183416942232282\\
86 0.0177977917785287\\
87 0.0176873310276924\\
88 0.0189094553574371\\
89 0.0191593276535546\\
90 0.0198398759412588\\
91 0.0209757201478261\\
92 0.0216117944690187\\
93 0.0222378079357813\\
94 0.023777090391183\\
95 0.0247733669837106\\
96 0.0243437823767133\\
97 0.0242138821244723\\
98 0.0240431870627262\\
99 0.0229736572195855\\
100 0.0230212465101702\\
};

\end{axis}
\end{tikzpicture}%
 \caption{Autocorrelation functions for the parameter $\theta$ (scale factor of transition covariance). Averages over 5 runs of PGAS without rejuvenation using $\Np=100$ particles (solid red) and $\Np=200$ particles (dashed blue), and with rejuvenation using $\Np=100$ particles (dotted green).}
 \label{fig:pgras:tracking_acf}
\end{figure}

Using this rejuvenation strategy roughly doubles the computation required to execute each iteration of \pgas. In simulations, the median factor of improvement in the probability of accepting an ancestor change at each time step is 2.4 (inter-quartile range 1.9--2.8). By contrast, simply doubling the number of particles results in a median factor of improvement of only 1.5 (inter-quartile range 1.4--1.6). The difference is even more pronounced in terms of autocorrelation, as shown in Figure~\ref{fig:pgras:tracking_acf}. (We also ran the
basic \pg algorithm, without ancestor sampling, but due to path degeneracy
the method did not converge and the results are therefore not reported here.)

\subsection{Extended target distribution}
The formal motivation for the validity of \pmcmc algorithms is based on an auxiliary variables argument.
More precisely, \citet{AndrieuDH:2010} introduce an \emph{extended target distribution}
which is defined on the space of all the random variables generated by the run of an \smc algorithm.
Let
\begin{align*}
  \xx_t &\eqdef \crange{x_t^1}{x_t^\Np}, &&\text{and} &   \aa_t &\eqdef \crange{a_t^1}{a_t^\Np},
\end{align*}
denote the particles and ancestor (resampling) indexes generated at time $t$, respectively.
We also write
\(
  \XX_t \eqdef \crange{\xx_1}{\xx_t}
\)
and
\(
  \AA_t \eqdef \crange{\aa_2}{\aa_t}.
\)
Furthermore, let $k$ be the index of one specific reference trajectory. To
make the particle indexes of the reference trajectory $\X_\T^k$ explicit we define recursively:
$b_\T = k$ and $b_t = a_{t+1}^{b_{t+1}}$ for $t < \T$. Hence, $b_t$ corresponds to the index
of the reference particle at time $t$, obtained by tracing the ancestry of $x_\T^k$.
If follows that $\X_\T^k = \prange{x_1^{b_1}}{x_\T^{b_\T}}$.

The extended target distribution for \pmcmc samplers is then given by (see~\cite{AndrieuDH:2010})
\begin{align}
  \label{eq:pgras:extended-target-def}
  \extpdf( \XX_\T, \AA_\T, k) \eqdef
  \frac{\ntDist{\theta}{\T}( \X_\T^k )}{\Np^\T}
  \Bigg\{ \prod_{\substack{i=1 \\ i\neq b_1 }}^\Np \r{\theta}{1}(x_1^i) \Bigg\} 
  \prod_{t = 2}^{\T} \Bigg\{ \prod_{\substack{i=1 \\ i\neq b_t }}^\Np
  \frac{w_{t-1}^{a_t^i}}{\sum_{\normind=1}^\Np w_{t-1}^\normind } \r{\theta}{t}(x_t^i \mid \X_{t-1}^{a_t^i})
  \Bigg\} .
\end{align}
A key property of this distribution is that it admits the original target distribution
$\ntDist{\theta}{\T}$
as a marginal. That is, if $(\XX_\T, \AA_\T, k)$ are distributed according to $\extpdf$,
then the marginal distribution of $\X_T^k$ is $\ntDist{\theta}{\T}$. This implies
that $\extpdf$ can be used in place of $\ntDist{\theta}{\T}$ in an \mcmc scheme; this is
the technique used by \pmcmc samplers.

\subsection{Partial collapsing and particle rejuvenation}\label{sec:pgras-pgras}
In particular, the \pgas algorithm that we reviewed in Section~\ref{sec:pgas}
corresponds to a partially collapsed Gibbs sampler for the extended target distribution $\extpdf$.
The complete Gibbs sweep corresponding to Algorithm~\ref{alg:pgas-original} is given in the appendix.
%
Here, however, we will focus on the ancestor sampling step~\eqref{eq:pgas:original_as}.
As mentioned in Remark~\ref{rem:as-important}, this step is very
useful for improving the mixing of the \pg algorithm.
However, as described above, for certain classes of models
the likelihood of the ``future'' reference path $\FX_t^\prime$
can be very low under alternative histories $\{ \X_{t-1}^i \}_{i=1}^\Np$.
The \pdf ratio in expression \eqref{eq:pgas:original_as} will thus cause the ancestor sampling distribution
to be highly concentrated on $i = b_{t-1}$, \ie, $\Prb(a_t^{b_t} = b_{t-1}) \approx 1$.
In particular, this is true for nearly degenerate \ssm{s}; recall model class \ma.
In fact, for truly degenerate models it may be that $\Prb(a_t^{b_t} = b_{t-1}) = 1$,
which implies that the ancestor sampling step has no effect 
and \pgas is reduced to the basic \pg scheme.

Observe that, while the \pgas algorithm attempts to update the \emph{ancestry} of the reference
particles $\X_T^\prime$, it does not update the values of the particles themselves.
This observation can be used to mitigate the aforementioned shortcoming of the algorithm,
as we will now illustrate.
The proposed modification is conceptually simple, but
its practical implications for improving the mixing of the \pgas algorithm
can be quite substantial for many models of interest.

The idea is to \emph{simultaneously} update the ancestor index $a_t^{b_t}$ together with a part of
the future reference trajectory $\FX_t^\prime$. This results in an increased flexibility of bridging
the future reference path with an alternative history, thereby increasing the probability of
changing its ancestry.  By ``a part of'', we here refere to any collection of random variables
$\UX_t \subseteq \FX_t$ (see Figure~\ref{fig:pgas:set-notation} for an illustration).  Typically,
the larger this subset is, the larger will the increased flexibility be.  However, this has to be
traded off with the difficulty of updating $\UX_t$, which can be substantial if $\UX_t$ is overly
high-dimensional.
%

For notational convenience,
let $\RX_t = \FX_t \setminus \UX_t$.
The AS step of the PGAS algorithm is then replaced by a step where we simulate $(a_t^{b_t}, \Xi_t)$
jointly from the conditional distribution on $\crange{1}{\Np} \times \rg{\Xi_t}$:
\begin{align}
  \label{eq:pgas:partial-collapse}
  \extpdf(a_t^{b_t}, \UX_t \mid \XX_{t-1}, \AA_{t-1}, \RX_t, \FB_{t}),
\end{align}
where $\FB_t \eqdef \prange{b_t}{b_\T}$.
Note that this is a so-called \emph{partially collapsed} Gibbs move, since not
all the non-simulated variables of the model are conditioned upon. Specifically, we
have excluded all the future particles and ancestor indexes, except for those corresponding to the
reference path. The justification for this is that we are sampling conceptually from the distribution
\begin{align}
  \extpdf(a_t^{b_t}, \UX_t, \{\FXX_{t}\setminus\FX_t\}, \{\FAA_t\setminus\FB_t\} \mid \XX_{t-1}, \AA_{t-1}, \RX_t, \FB_{t}),
\end{align}
which is a standard Gibbs update for the extended target distribution \eqref{eq:pgras:extended-target-def}. However, no consecutive operation
will depend on the variables $\FXX_{t}\setminus\FX_t$ and $\FAA_t\setminus\FB_t$, which
has the implication that these variables need not be generated at all; see \cite{DykP:2008}.

Following \cite{LindstenJS:2014} we obtain an expression
for the conditional distribution \eqref{eq:pgas:partial-collapse} which much resembles
the original ancestor sampling distribution \eqref{eq:pgas:original_as}, namely,
\begin{align}
  \label{eq:pgas:partial-collapse-explicit}
  \extpdf(a_t^{b_t}, \UX_t \mid \XX_{t-1}, \AA_{t-1}, \RX_t^\prime, \FB_{t}) 
  \propto
  w_{t-1}^{a_t^{b_t}} \frac{\tDist{\theta}{\T}( \X_{t-1}^{a_t^{b_t}} \cup \UX_t \cup \RX_t^\prime  )}{ \tDist{\theta}{t-1} (\X_{t-1}^{a_t^{b_t}}) }.
\end{align}
Note, however, that this is a distribution on $\crange{1}{\Np} \times \rg{\Xi_t}$, and simulating
from this distribution allows us to update the ancestor index $a_t^{b_t}$ jointly with a
part of the reference trajectory $\UX_t$.
Additionally, at time $t = 1$ we can update $\UX_1$ by simulating from the conditional distribution
\begin{align}
  \label{eq:pgas:partial-collapse-t1}
  \extpdf(\UX_1 \mid \RX_1^\prime, \FB_1) \propto
  \tDist{\theta}{\T}( \UX_1 \cup \RX_1^\prime ).  
\end{align}
In most cases, exact simulation from \eqref{eq:pgas:partial-collapse-explicit} or \eqref{eq:pgas:partial-collapse-t1}
is not possible. However, this issue can be dealt with by instead simulating from some \mcmc kernels leaving these
distributions invariant, resulting in a standard combination of \mcmc samplers, see \eg \cite{Tierney:1994}.
Hence, let $K_t$ denote a Markov kernel on $\crange{1}{\Np} \times \rg{\Xi_t}$
which leaves the distribution~\eqref{eq:pgas:partial-collapse-explicit} invariant
(sampling from \eqref{eq:pgas:partial-collapse-t1} at time $t=1$ follows analogously).
The proposed modified \pgas method is then given by Algorithm~\ref{alg:pgras}.

\begin{algorithm}
  \caption{\pgas with particle rejuvenation}
  \label{alg:pgras}
  \begin{algorithmic}[1]
    \REQUIRE Reference trajectory $\X_\T^\prime \in \setX^\T$.
    \STATE Simulate $\UX_1^{\star} \sim K_1(\UX_1^\prime, \cdot) $ and update $\X_\T^\prime$ accordingly:
    $\X_\T^\prime \leftarrow \{ \X_T^\prime \setminus \UX_1^\prime \} \cup \UX_1^{\star}$.%
    \STATE Set $x_1^{\Np} = x_1^\prime$.
    \STATE Draw $x_1^i \sim \r{\theta}{1}(\cdot)$ for $i = \range{1}{\Np-1}$.
    \STATE Set $w_1^i = \tDist{\theta}{1}(x_1^i)/\r{\theta}{1}(x_1^i)$ for $i = \range{1}{\Np}$.
    \FOR{$t = 2$ \TO $\T$}
    \STATE Simulate $( a_t^i, x_t^i)$ as in (\ref{eq:bkg:resampling}, \ref{eq:bkg:propagation}) for $i =\range{1}{\Np-1}$.
    \STATE Simulate $(a_t^\Np, \UX_t^{\star}) \sim K_t( (\Np, \UX_t^{\prime}), \cdot)$ and update
    $\X_\T^\prime$ accordingly: $\X_\T^\prime \leftarrow \{ \X_T^\prime \setminus \UX_t^\prime \} \cup \UX_t^{\star}$.%
    \STATE Set $x_t^{\Np} = x_{t}^\prime$.
    \STATE Set $\X_{t}^i = ( \X_{t-1}^{a_t^i}, x_t^i )$ for $i = \range{1}{\Np}$.
    \STATE Set $w_t^i = \wf{\theta}{t}(\X_{t}^{i})$ for $i = \range{1}{\Np}$.
    \ENDFOR
    \STATE Draw $k$ with $\Prb(k = i) \propto w_\T^i$.
    \RETURN $\X_{\T}^\star = \X_{\T}^k$.
  \end{algorithmic}
\end{algorithm}

\begin{remark}
  The previous particle rejuvenation strategies proposed independently by us \cite{BunchLS:2015}
  and \citet{CarterMK:2014} correspond to the special case obtained by setting $\UX_t = x_t$.
  However, as we shall see in Sections~\ref{sec:degen}~and~\ref{sec:abc}, this is insufficient
  in many cases. In particular, to address the challenges associated with some degenerate models \ma
  and models with intractable transitions \mb, we need additional flexibility in selecting~$\UX_t$.
  Furthermore, a difference between the current derivation and the one presented by \citet{CarterMK:2014}
  is that they do not make use of the technique of \emph{partial collapsing}. As a consequence,
  they are forced to re-define the extended target distribution~\eqref{eq:pgras:extended-target-def},
  resulting in (unnecessary) modifications of the \smc scheme and it implies that their approach is only applicable
  when using an explicit backward pass (as in PGBS).
\end{remark}

\begin{remark}
  While ergodicity of the kernels $K_t$ is of practical importance in order to obtain
  a large performance improvement from the 'ancestor sampling \& particle rejuvenation' strategy,
  it is not needed to guarantee ergodicity
  of the overall sampling scheme. In particular, if $K_t( (a_t^{b_t}, \UX_t), \cdot) = \delta_{(a_t^{b_t}, \UX_t)} (\cdot)$,
  the proposed method reduces to the original \pg algorithm by \cite{AndrieuDH:2010}
  which is known to be uniformly geometrically ergodic under weak assumptions \cite{LindstenDM:2015,AndrieuLV:2013}.
\end{remark}

Below, we present two specific techniques for designing the kernels $K_t$ that can be useful in the present context.
\begin{paragraph}{Metropolis-Hastings (MH)}
We can target \eqref{eq:pgas:partial-collapse-explicit} using MH. From current values $(a_t^\prime, \UX_t^\prime)$, we can propose new values $(a_t^{\star}, \UX_t^{\star})$ by drawing from,
\begin{align}
  \label{eq:pgras:mh-proposal}
  \frac{ \ppw{t-1}{a_t} }{ \sum_{\normind=1}^\Np \ppw{t-1}{\normind} } \pd{t}( \UX_t \mid \X_{t-1}^{a_t} , \UX_t^\prime, \RX_t^\prime ) ,
\end{align}
where $\{\ppw{t}{i}\}_{i=1}^\Np$ is a set of proposal weights for the ancestor index and $\pd{t}$ is a proposal density
for rejuvenating the reference particles $\UX_t^\prime$. The resulting acceptance probability is then,
\begin{multline}
  \alpha\left\{ (a_t^\prime, \UX_t^\prime) \rightarrow (a_t^{\star}, \UX_t^{\star}) \right\}\\
  = \min\Bigg\{1, 
    \frac{ \ppw{t-1}{a_t^\prime} \pd{t}( \UX_t^\prime \mid \X_{t-1}^{a_t^\prime} , \UX_t^{\star}, \RX_t^\prime ) }{  \ppw{t-1}{a_t^{\star}} \pd{t}( \UX_t^{\star} \mid \X_{t-1}^{a_t^{\star}} , \UX_t^{\prime}, \RX_t^\prime )}
  \frac{ w_{t-1}^{a_t^{\star}} \tDist{\theta}{\T}( \X_{t-1}^{a_t^{\star}} \cup \UX_t^{\star} \cup \RX_t^\prime  )}{ w_{t-1}^{a_t^{\prime}} \tDist{\theta}{\T}( \X_{t-1}^{a_t^{\prime}} \cup \UX_t^\prime \cup \RX_t^\prime  )}
\frac{ \tDist{\theta}{t-1} (\X_{t-1}^{a_t^{\prime}}) }{ \tDist{\theta}{t-1} (\X_{t-1}^{a_t^{\star}}) }
  \Bigg\}.
\end{multline}
%
\end{paragraph}

\begin{paragraph}{Conditional importance sampling}
Given that we are working within the \pg framework, a more natural
approach might be to use a conditional importance sampling (\cis) Markov kernel.
This can be viewed simply as an instance of the \pg kernel applied to a single time step.
Consider an importance sampling proposal distribution for $(a_t, \UX_t)$ (\cf~\eqref{eq:pgras:mh-proposal}),
\begin{align}
  \label{eq:pgras:cis-proposal}
  \frac{ \ppw{t-1}{a_t} }{ \sum_{\normind=1}^\Np \ppw{t-1}{\normind} } \pdd{t}( \UX_t \mid \X_{t-1}^{a_t} , \RX_t^\prime ) .
\end{align}
Given the current values $(a_t^\prime, \UX_t^\prime)$, a Markov kernel with
\eqref{eq:pgas:partial-collapse-explicit} as its stationary distribution can be constructed
as in Algorithm~\ref{alg:cis}. The validity of this approach follows as a special case of the
derivation of the \pg kernel \cite{AndrieuDH:2010}.

\begin{algorithm}
  \caption{Conditional importance sampling}
  \label{alg:cis}
  \begin{algorithmic}[1]
    \REQUIRE Current state $(a_t^\prime, \UX_t^\prime)$.
    \STATE Set $\widecheck a_t^\Np = a_t^\prime$ and $\widecheck \UX_t^{\Np} = \UX_t^\prime$.
    \STATE Draw $(\widecheck a_t^i, \widecheck \UX_t^i)$ from 
    \eqref{eq:pgras:cis-proposal} for $i = \range{1}{\Np-1}$.
    \STATE \label{step:cis:weights} Set $$\widecheck w_{t-1}^i = \frac{w_{t-1}^{\widecheck a_t^{i}} \tDist{\theta}{\T}( \X_{t-1}^{\widecheck a_t^{i}} \cup \widecheck \UX_t^{i} \cup \RX_t^\prime )}{\ppw{t-1}{\widecheck a_t^i} \pdd{t}( \widecheck\UX_t^i \mid \X_{t-1}^{\widecheck a_t^i} , \RX_t^\prime ) \tDist{\theta}{t-1} (\X_{t-1}^{\widecheck a_t^{i}}) } $$ for ${i = \range{1}{\Np}}$.
    \STATE Draw $\ell$ with $\Prb(\ell = i) \propto \widecheck w_{t-1}^i$.
    \RETURN $(a_t^{\star}, \UX_t^{\star}) = (\widecheck a_t^\ell, \widecheck\UX_t^\ell)$.
  \end{algorithmic}
\end{algorithm}

\end{paragraph}

\begin{remark}
  We can also define the kernel $K_t$ to be composed of $m$, say, iterates of the MH or the CIS kernel
  to improve its mixing speed (at the cost of an $m$-fold increase in the computational cost of simulating
  from $K_t$). Indeed, any standard combination of \mcmc kernels (see, \eg, \cite{Tierney:1994}) targeting
  \eqref{eq:pgas:partial-collapse-explicit} will result in a valid definition of $K_t$.
\end{remark}

\subsection{Convergence properties}
Existing convergence analysis for particle Gibbs algorithms \cite{LindstenDM:2015,AndrieuLV:2013,ChopinS:2014}
can be extended also to the proposed modified \pgas procedure of Algorithm~\ref{alg:pgras}. Here, we restate the uniform ergodicity result
for the \pgas algorithm presented by \cite{LindstenJS:2014}, adopted to the current settings.
We write $\supnorm{\cdot}$ and 
$D_{\TV}$ for the supremum norm and the total variation distance, respectively.

\begin{theorem} \label{thm:pgas:uniform}
  Assume that there exists a constant $\ergokappa < \infty$ such that
  $\supnorm{ \wf{\theta}{t} } \leq \ergokappa$ for any $t \in \crange{1}{\T}$.
  Then, for any  $\Np\geq2$ there exist constants $\ergoR < \infty$ and $\ergorho \in [0,1)$ such that
  \begin{align*}
    D_{\TV} (\operatorname{Law}(\X_\T[k]), \ntDist{\theta}{\T}) 
    \leq \ergoR \ergorho^{n},
    &&\forall \X_{\T}' \in \setX^\T,
  \end{align*}
  where the Markov chain $\{ \X_\T[k] \}_{k\geq 0}$ is generated by iterating Algorithm~\ref{alg:pgras} with
  initial state $\X_T[0] = \X_\T'$.
\end{theorem}

The proof follows analogously to the proof of \cite[Theorem~3]{LindstenJS:2014}
and is omitted for brevity.

\section{Degenerate and nearly degenerate models}\label{sec:degen}
The method proposed in Algorithm~\ref{alg:pgras} can be used for a general
sequence of target distributions $\{\tDist_t(\X_t)\}_{t=1}^\T$.
We now turn our attention explicitly to (nearly) degenerate \ssm{s} as described
in \ma and discuss how Algorithm~\ref{alg:pgras} can be used
for these models.

\subsection{Ancestor sampling for nearly degenerate models}
Consider again inference for the \ssm given by \eqref{eq:ssm},
with the unnormalised target density $\tDist{\theta}{t}(\X_{t}) = p(\X_{t}, \Y_{t})$.
We follow the convention used in Algorithms~\ref{alg:pgas-original}~and~\ref{alg:pgras},
that the reference particle is always placed on the $\Np$th position.
It follows that the ancestor sampling distribution \eqref{eq:pgas:original_as} is given by
\begin{align}
  \label{eq:degen:as_step}
  \Prb(a_t^\Np = i) &=  \frac{ w_{t-1}^i f(x_{t}^\prime \mid x_{t-1}^i) }{ \sum_{\normind=1}^\Np w_{t-1}^\normind f(x_{t}^\prime \mid x_{t-1}^\normind) }, &i = \range{1}{\Np}.
\end{align}
If the state process noise is small, \ie the transition density $f(\cdot)$ is nearly degenerate, then
this probability distribution can be highly concentrated on $i=\Np$, effectively removing the effect of ancestor
sampling; we experienced this effect in the motivating example in Section~\ref{sec:pgras:motivating}.

To cope with this issue, one option is to make a partial collapse over a subset of the future state variables.
That is, we let $\UX_t = \prange{x_t}{x_{\kappa_t}}$ where $\kappa_t = \min\{\T, t+\ell-1\}$ for some fixed length $\ell$.
It follows that the ratio of the unnormalised target densities appearing in \eqref{eq:pgas:partial-collapse-explicit}
can be written as
\begin{align}
  \frac{ \tDist{\theta}{T}(\X_{\T}) }{\tDist{\theta}{t-1}(\X_{t-1})} = p(\FX_{t}, \FY_{t} \mid x_{t-1})
  \propto  f(x_{\kappa_t+1} \mid x_{\kappa_t}) \left\{ \prod_{s = t}^{\kappa_t} f(x_s \mid x_{s-1}) g(y_s \mid x_s) \right\}.
\end{align}
Hence, the target distribution for the modified ancestor sampling step,
with particle rejuvenation, is defined on $\crange{1}{N}\times \setX^{\kappa_t-t+1}$
and the corresponding \pdf of $(a_t, \UX_t)$ is proportional to
\begin{align}
  w_{t-1}^{a_t}  f(x_{\kappa_t+1}^\prime \mid x_{\kappa_t}) 
  \left\{ \prod_{s = t+1}^{\kappa_t} f(x_s \mid x_{s-1}) g(y_s \mid x_s) \right\}
  f(x_t \mid x_{t-1}^{a_t}) g(y_t \mid x_t).
\end{align}
Simulating from this distribution can be done, \eg, by using one of the \mcmc kernels introduced in Section~\ref{sec:pgras-pgras}. The benefit of doing this is that by rejuvenating the reference
trajectory over the variables $\prange{x_t}{x_{\kappa_t}}$ we are able to bridge between
$x_{t-1}^i$ and $x_{\kappa_t+1}^\prime$, thereby increasing the probability of changing the ancestry
for the reference path.

We used this approach with $\ell=1$ and a CIS Markov kernel for state rejuvenation for the target tracking model
in Section~\ref{sec:pgras:motivating}.
Below, we present two other example applications where the aforementioned technique can be useful.

\subsection{Example: Euler-Maruyama discretisation of SDEs}
Consider a continuous-time state space model with hidden state $\{\cX(\ct)\}_{\ct\geq 0}$,
represented by the following stochastic differential equation (SDE)
\begin{align}
  d\cX_\ct = \mu(\cX_\ct)d\ct + \sigma(\cX_\ct)dW_\ct,
\end{align}
where $W_{\ct}$ denotes a Wiener process. The process is observed indirectly through the observations
$\prange{y_1}{y_\T}$, obtained at time points $\prange{\ct_1}{\ct_\T}$, where
$y_t \sim g(y_t \mid \cX_{\ct_t})$.  A simple approach to enable inference in this model is to
consider a time discretisation of the continuous process, using an Euler-Maruyama scheme, after
which standard discrete-time inference techniques can be used.
For simplicity, assume that the observations are
equidistant, $\Delta\ct \eqdef \ct_t - \ct_{t-1}$, and that we sample
the process $m$ times for each observation.

Let the discrete-time state at time $\ct_t$
consist of $\cX_{\ct_t}$, as well as the $m-1$ intermediate states, \ie
\begin{align}
  x_t \eqdef
  \begin{pmatrix}
    \iX_{t,1}^\+ & \cdots & \iX_{t,m-1}^\+ &  \cX_{\ct_t}^\+
  \end{pmatrix}^\+
\end{align}
where $\iX_{t,j} =  \cX_{\ct_{t-1}+j\Delta\ct}^\+$ for $j=\range{1}{m-1}$.
When using PGAS for this model, a problem is that, while increasing $m$ makes
the discretisation more accurate, it will also make the transition kernel of the latent process more degenerate.

However, this issue can be mitigated by rejuvenating the intermediate state variables,
\ie the state variables in between observation time points.
Hence, we set
\begin{align}
  \UX_t = 
  \begin{pmatrix}
    \iX_{t,1}^\+ & \cdots & \iX_{t,m-1}^\+
  \end{pmatrix}^\+.
\end{align}
Similarly to above, it follows that the ratio of unnormalised target densities is given by
\begin{align}
  \frac{ \tDist{\theta}{T}(\X_{\T}) }{\tDist{\theta}{t-1}(\X_{t-1})} &\propto f(x_{t} \mid x_{t-1}) \\
  &\propto p(\cX_{\ct_t} \mid  \iX_{t,m-1} ) \left\{ \prod_{j=2}^{m-1} p( \iX_{t,j} \mid  \iX_{t,j-1}) \right\} p( \iX_{t,1} \mid \cX_{\ct_{t-1}}) ,
\end{align}
where, by the Euler-Maruyama discretisation,
\begin{align}
  p( \cX_{\ct+\Delta\ct} \mid  \cX_{\ct}) \approx \N( \cX_\ct + \mu( \cX_\ct)\Delta\ct, \sigma^2(\cX_\ct)\Delta\ct).
\end{align}
Hence, the unnormalised target \pdf in \eqref{eq:pgas:partial-collapse-explicit} is given by
\begin{align}
  w_{t-1}^{a_t} p(\cX_{\ct_t}^\prime \mid  \iX_{t,m-1} ) \left\{ \prod_{j=2}^{m-1} p( \iX_{t,j} \mid \iX_{t,j-1}) \right\}  p( \iX_{t,1} \mid \cX_{\ct_{t-1}}^{a_t}).
\end{align}
To obtain an efficient \mcmc proposal distribution for this \pdf, we can use
one of the methods proposed by \cite{DurhamG:2002} which are based on a tractable diffusion bridges
between $\cX_{\ct_{t-1}}^{a_t}$ and~$\cX_{\ct_t}^\prime$.

\subsection{Example: Degenerate Gaussian transition}\label{sec:degen:gaussian}
Consider a model with a Gaussian transition, but a possibly nonlinear/non-Gaussian
observation
\begin{subequations}
  \label{eq:degen-lg}
  \begin{align}
  \label{eq:degen-lg-a}
    x_{t+1} &= Ax_t + F \drv_{t+1}, \\
    y_t &\sim g(y_t \mid x_t),
  \end{align}
\end{subequations}
with $v_t \sim \N(0,I_d)$.
Furthermore, assume that $\dim(x_t) = n$ and that $\rank(F) < n$. This implies
that the transition kernel of the linear Gaussian state process is degenerate.
Models on this form are common in certain application areas, \eg, navigation
and tracking; see \cite{GustafssonGBFJKN:2002} for several examples.

In this case, the ancestor sampling step is even more problematic than
for the previous example, since the transition is truly degenerate.
Indeed, if we use the ancestor sampling distribution from \eqref{eq:pgas:original_as},%
\footnote{Recall that we use density notation also for the degenerate kernel.}
the probability of selecting $x_{t-1}^i$ as the ancestor of $x_t^\prime$
will be zero, unless $x_t^\prime - Ax_{t-1}^i$ is in the column space of $F$.
However, in general, this will almost surely not be the case, except for $i=\Np$.
Thus, the distribution \eqref{eq:pgas:original_as} puts all probability mass on $i=\Np$,
resulting in zero probability of changing the ancestry of the reference trajectory.
In fact, it is not only for \pgas the degeneracy of the transition kernel is problematic.
As discussed in \cite[Section~4.6]{LindstenS:2013}, any conventional \smc-based
forward-backward smoother will be inapplicable for the model \eqref{eq:degen-lg}
due to the degeneracy of the backward kernel.

However, by collapsing over intermediate state variables, this problem can be circumvented.
We assume that the pair $(A,F)$ in \eqref{eq:degen-lg} are controllable (see \eg, \cite{KailathSH:2000}
for a definition).
Informally, this means that any state in the state space is reachable from any other state,
\ie for any $(x,x^\prime)\in\setX^2$ there exists an integer $\ell$ and a noise realisation $\drv_{t:t+\ell}$
which takes the system from $x_{t-1} = x$ at time $t-1$ to $x_{t+\ell}=x^\prime$ at time $t+\ell$.
Now, let $\UX_t = \prange{x_{t}}{x_{\kappa_t}}$ with $\kappa_t = \min\{\T, t+\ell-1\}$ and
where the length $\ell$ is chosen as any integer (\eg, the smallest) such that the matrix
\begin{align*}
  C_\ell =
  \begin{bmatrix}
    F & AF & \cdots & A^{\ell}F
  \end{bmatrix}
\end{align*}
is of rank $n$ (the existence of such an integer is guaranteed by the controllability assumption).

Assuming $\kappa_t < \T$ (the case $\kappa_t = \T$ follows analogously)
we have that the unnormalised target \pdf in \eqref{eq:pgas:partial-collapse-explicit} is given by
\begin{align}
  \label{eq:degen:target}
  w_{t-1}^{a_t} \left\{ \prod_{s = t}^{t+\ell-1} g(y_s \mid x_s) \right\} p(x_{t+\ell}^\prime, \UX_t \mid x_{t-1}^{a_t}),
\end{align}
where $p(x_{t+\ell}^\prime, \UX_t \mid x_{t-1}^{a_t})$ corresponds to the prior distribution
under the linear Gaussian dynamics \eqref{eq:degen-lg-a}
of the
state sequence $\UX_t$ and the end-point $x_{t+\ell}^\prime$, conditionally on the starting point $x_{t-1}^{a_t}$.
Even though this distribution is degenerate for the model \eqref{eq:degen-lg},
our choice of~$\ell$ ensures that for any $(x_{t-1}, x_{t+\ell})\in\setX^2$,
there exists a $\UX_t \in \setX^{\kappa_t-t+1}$ in the support of the distribution.

In particular, the conditional distribution $p(\UX_t \mid x_{t-1}^{a_t}, x_{t+\ell}^\prime)$ is a (degenerate)
Gaussian distribution which it is possible to sample from. For instance, this can be done by running
a Kalman filter/backward simulator \cite{CarterK:1994,Fruhwirth-Schnatter:1994}
for time steps $\range{t}{t+\ell-1}$ for the state process \eqref{eq:degen-lg-a},
with $x_{t+\ell}^\prime = Ax_{t+\ell-1} + Fv_{t+\ell}$ acting as an ``observation'' at
the final time step (\ie, we condition on the reference state $x_{t+\ell}^\prime$ via a standard
measurement update of the Kalman filter).
Consequently, it is possible to use this as a proposal distribution in the MH kernel \eqref{eq:pgras:mh-proposal}
or in the \cis kernel \eqref{eq:pgras:cis-proposal} to simulate from \eqref{eq:degen:target}. Indeed,
in taking the ratio between the target \eqref{eq:degen:target} and the proposal $p(\UX_t \mid x_{t-1}, x_{t+\ell})$
we have $ p(x_{t+\ell}, \UX_t \mid x_{t-1}) / p(\UX_t \mid x_{t-1}, x_{t+\ell}) =
p(x_{t+\ell} \mid x_{t-1})$ which is a well-defined Gaussian density for any $(x_{t-1}, x_{t+\ell})\in\setX^2$.
Specifically,
\begin{align}
  \label{eq:degen:likelihood-future-state}
  p(x_{t+\ell} \mid x_{t-1}) = \N(x_{t+\ell} \mid A^{\ell+1} x_{t-1}, C_\ell C_\ell^\+),
\end{align}
which is non-degenerate under the controllability condition.

\section{Near-degenerate approximations of intractable transitions}\label{sec:abc}
Another class of \ssm{s} which poses large inferential challenges are models with
intractable transition density functions as explained under label \mb.
Hence, consider an \ssm on the form \eqref{eq:ssm} and
assume that the transition density function $f(\cdot)$ is a regular, non-degenerate \pdf which
it is possible to simulate from, but which is not available for evaluation in closed form.
A problem with these models is that the backward kernel \eqref{eq:bkg:bwd-kernel} is also intractable,
 essentially ruling out any forward-backward-based inference technique.
%
In fact, one of the main merits of the \pmcmc samplers derived in \cite{AndrieuDH:2010}
is that, in their most basic implementations, they only require forward simulation of the system
dynamics. These methods---specifically, \pg and the particle independent Metropolis-Hastings (\pimh) sampler---can thus be readily used for inference in models with intractable transitions.
However, these methods (\pg and \pimh)
are liable to poor mixing unless a large number of particles are used in the underlying \smc samplers
(see, \eg, \cite{LindstenJS:2014}).
Intuitively, the reason for this is that we require the \smc sampler to generate approximate
draws from the full \emph{joint smoothing distribution}, which is difficult using only forward-simulation
due to path space degeneracy.

As has been demonstrated (here and in the previous literature, \eg, \cite{LindstenJS:2014,LindstenS:2013}),
the \pgas sampler will in many cases enjoy much better mixing than \pg and \pimh,
in particular when using few particles $\Np$ relative to the number of observations $\T$.
However, the \pgas sampler is not directly applicable to
models with intractable transitions. Indeed, to simulate from the
ancestor sampling distribution \eqref{eq:degen:as_step} it is necessary to evaluate
the (intractable) transition \pdf $f(\cdot)$. In this section, we propose one way
to address this limitation. The idea is to approximate, as detailed below, the intractable transition \pdf
with a nearly degenerate transition. Using the proposed
particle rejuvenation procedure, much in the same way as discussed in Section~\ref{sec:degen},
we can then enable \pgas for this challenging class of \ssm{s}.



The proposed method is essentially a variant of the approximate Bayesian computation (\abc)
technique \cite{BeaumontZB:2002,TavareBGD:1997}.
However, while \abc is typically used for inference in models with intractable \emph{likelihoods}
(see, \eg, \cite{DeanSJP:2014,McKinleyCD:2009} for \smc implementations), we use it here
to address the issue of intractable transitions. The idea is based on the realisation
that 
simulating from the transition \pdf $f(\cdot)$, which is assumed to be feasible, is always done by generating some
``driving noise variable'' $\drv_t$, say, which is then propagated through some function $\Gamma(\cdot)$ (note that this function can be implicitly defined by a computer program or simulation-based software). By explicitly introducing
these noise variables, we can thus rewrite the original model on the equivalent form,%
\begin{subequations}
  \label{eq:abc:degen_ssm}
  \begin{align}
    \drv_t&\sim p_v(\drv_t), \\
    x_t&= \Gamma(x_{t-1},  \drv_t), \\
    y_t&\sim g(y_t \mid x_t).
  \end{align}
\end{subequations}
Note that $x_t$ here is given by a deterministic mapping of $x_{t-1}$ and $\drv_t$.
Consequently, the transition function for the \emph{joint state} $(\drv_t, x_t)$ is given by
\begin{align}
  \label{eq:abc:degen-kernel}
  p(\drv_t, x_t \mid \drv_{t-1}, x_{t-1})
 = p_v(\drv_t) \delta_{\Gamma(x_{t-1},  \drv_t)}(x_t),
\end{align}
which is a degenerate transition kernel due to the Dirac measure on $x_t$.

\begin{remark}
The reformulation given by \eqref{eq:abc:degen_ssm} can be seen as
transforming the difficulty of having an intractable transition, to that of having a degenerate one.
Now, if we marginalise over $x_{1:t}$ (which is straightforward since $x_{1:t}$ is deterministically given by $\drv_{1:t}$)
we obtain a model specified only in the noise variables~$\{ \drv_t \}_{t\geq 1}$. This approach has previously
been used by \citet{MurrayJP:2013} in the context of auxiliary \smc sampling
and for particle marginal Metropolis-Hastings. It was also used by
\citet{LindstenJS:2014} to enable inference by \pgas in a model with an intractable transition.
The problem with that approach, however, is that the marginalisation of the $x_t$-process
introduces a non-Markovian dependence in the observation likelihood, resulting in a $\T^2$ computational complexity
for \pgas (see \cite{LindstenJS:2014} for details).
\end{remark}

Simply rewriting the model as in \eqref{eq:abc:degen_ssm} does not solve the problem, since,
as discussed above, the degenerate transition kernel is problematic when using \pgas.
However, by making use of an \abc approach this issue can be addressed.
Specifically, we make use of a near-degenerate approximation of \eqref{eq:abc:degen-kernel}.
In the ancestor sampling step of the algorithm we replace the point-mass distribution by some (for instance, Gaussian) kernel $\ABCKernel{\epsilon}: \setX^2 \mapsto \nonnegatives$
centered on $\Gamma(x_{t-1},  \drv_t)$:
\begin{align}
  \label{eq:abc:kernel-approx}
  \delta_{\Gamma(x_{t-1}, \drv_t)}(x_t) 
  \approx \ABCKernel{\epsilon}( \Gamma(x_{t-1},  \drv_t), x_t),
\end{align}
where $\epsilon$ controls the band-width of the kernel (and thus the approximation error).
Next, to deal with the near-degeneracy of the approximation (for small~$\epsilon$)
we select $\UX_t = v_t$ to be rejuvenated, which results in a joint target for $(a_t,\drv_t)$, as in \eqref{eq:pgas:partial-collapse-explicit}, proportional to
\begin{align}
  w_{t-1}^{a_t} p_v(\drv_t) \ABCKernel{\epsilon}( \Gamma(x_{t-1}^{a_t},  \drv_t), x_t^\prime).
\end{align}
The connection to \abc is perhaps most easily seen if we make use of the \cis kernel
given in Algorithm~\ref{alg:cis} for simulating from this distribution. Let
$\pdd{t}( \drv_t \mid \X_{t-1} , \RX_t^\prime ) = p_v(\drv_t)$
be the proposal distribution for the noise variable
(in many cases this is likely to be the only sensible choice) and let $\ppw{t-1}{a_t} = w_{t-1}^{a_t}$ in \eqref{eq:pgras:cis-proposal}. We then obtain the following ancestor sampling
procedure, using the convention $a_t^\prime = N$ (\cf Algorithm~\ref{alg:cis}):
\begin{itemize}
\item For $i = \range{1}{\Np-1}$:
  \begin{itemize}
  \item Simulate $\widecheck a_t^i$ with $\Prb(\widecheck a_t^i = j) \propto w_{t-1}^j$. 
  \item Simulate a noise realisation $\widecheck \drv_t^i\sim p_v(\cdot)$ and set $\widecheck x_t^i = \Gamma(x_{t-1}^{\widecheck a_{t}^i}, \widecheck \drv_t^i)$.
  \item Compute $\widecheck w_{t-1}^i = \ABCKernel{\epsilon}(\widecheck x_t^i , x_t^\prime)$.
  \end{itemize}
\item Compute $\widecheck w_{t-1}^\Np = \ABCKernel{\epsilon}(x_t^\prime , x_t^\prime)$.
\item Simulate $\ell$ with $\Prb(\ell = i) \propto \widecheck w_{t-1}^i$, $i = \range{1}{\Np}$.
\item If $\ell < \Np$, return $a_t^\star = \widecheck a_t^\ell$, otherwise return $a_t^\star = \Np$.
\end{itemize}
In the above, we have assumed that the kernel approximation \eqref{eq:abc:kernel-approx}
is used only in the ancestor sampling step of the algorithm (\ie, in the forward simulation
of particles we use the original model \eqref{eq:abc:degen_ssm}).
An interesting implication of this is that there is no need to explicitly keep track of the $\drv_t$-variables.
Indeed, 
the CIS procedure outlined above can be expressed in words as follows: \emph{(i)} Generate
an independent set of $N-1$ resampled particles at time $t-1$ and, for each one, simulate the system dynamics
forward to obtain $\{\widecheck x_t^i \}_{i=1}^{N-1}$, \emph{(ii)} set the final particle according to the conditioning
$\widecheck x_t^\Np = x_t^\prime$, and \emph{(iii)} simulate a new ancestor for $x_t^\prime$ based on
the closeness of $\{\widecheck x_t^i\}_{i=1}^\Np$ to $x_t^\prime$, as measured by the kernel $\ABCKernel{\epsilon}$.

\begin{remark}
  In some cases, for instance if $\setX$ is high-dimensional, it can be beneficial to define
  the kernel $\ABCKernel{\epsilon}$ in \eqref{eq:abc:kernel-approx} on some summary statistic $S : x_t \mapsto S(x_t)$,
  rather than on the state variable itself; see, \eg, \cite{FearnheadP:2012} for details.
\end{remark}

\section{Numerical illustration}
In this section we illustrate the particle rejuvenation strategy for \pgas on two examples.
We have already seen the merits of the approach when compared to standard \pgas (and \pg) on a nearly degenerate
target tracking model in Section~\ref{sec:pgras:motivating}. Hence, here we consider two alternative model classes,
first a model with a linear Gaussian degenerate transition as discussed in Section~\ref{sec:degen:gaussian}
and then a model with an intractable transition as discussed in Section~\ref{sec:abc}.

\subsection{Degenerate transition model}

Autoregressive models are widely used to model stochastic processes \cite{Godsill1998}. They may be written in state space form as a degenerate Gaussian transition model,
\begin{equation}
 x_{t+1} =
   \underbrace{\begin{bmatrix}
     \alpha_1 & \alpha_2 & \dots & \alpha_{n-1} & \alpha_n \\
     1        & 0        & \dots & 0 & 0        \\
     0        & 1        & \dots & 0 & 0        \\
     \vdots   & \vdots   & \ddots& \vdots & \vdots        \\
     0        & 0        & \dots & 1 & 0
   \end{bmatrix}}_{=A}
 x_{t} +
   \underbrace{\begin{bmatrix}
    \sigma_v \\
    0 \\
    0 \\
    \vdots \\
    0
   \end{bmatrix}}_{=F}
   v_{t+1}  ,
\end{equation}
where $\alpha =
\begin{pmatrix}
  \alpha_1 & \cdots & \alpha_n
\end{pmatrix}^\+
$ is a vector of regression parameters, $\{v_t\}_{t\geq 1}$ is (scalar) white Gaussian noise, and $\sigma_v$ is the process noise standard deviation.
In this example, we model the latent state as an autoregressive process of order $n=5$. 
We simulate the system for $\T = 500$ time steps using $\alpha =
\begin{pmatrix}
  0.9 & -0.8 & 0.7 & -0.6 & 0.5
\end{pmatrix}^\+$ and $\sigma_v = 1$.
Each observation is a noisy, saturated measurement of the first component of $x_t$, modelled as,
\begin{align}
  y_t &= \beta^{-1} \text{tanh}(\beta x_{1,t}) + \sigma_e e_t,
\end{align}
where $e_t$ is $t$-distributed with $\nu = 3$ degrees of freedom. We set $\beta = 0.5$ and $\sigma_e = 0.5$.

Since $\rank ( FF^\+) = 1 < n$, the transition kernel is degenerate. Consequently,
standard ancestor sampling is ineffective, and \pgas without rejuvenation is equivalent to basic \pg.
However, by collapsing over states $\UX_t = \prange{x_t}{x_{t+\ell-1}}$ using $\ell \ge 4$, it is possible to update the particle ancestry.
We use the \cis Markov kernel with new ancestor indexes sampled proportional to the filter weights, and state sequences sampled according to $p(\UX_t \mid x_{t-1}, x_{t+\ell})$. The resulting CIS weights (see Algorithm~\ref{alg:cis}, Step~\ref{step:cis:weights}) are then,
\begin{align}
  \widecheck w_{t-1}^i = \left\{ \prod_{s = t}^{t+\ell-1} g(y_s \mid \widecheck x_s^i) \right\} p(x_{t+\ell}^\prime \mid x_{t-1}^{\widecheck a_t^{i}}),
\end{align}
where $p(x_{t+\ell} \mid x_{t-1})$ is a well-defined Gaussian density given by \eqref{eq:degen:likelihood-future-state}.

\begin{figure}[ptb]
  \centering
  \includegraphics[width=0.7\linewidth]{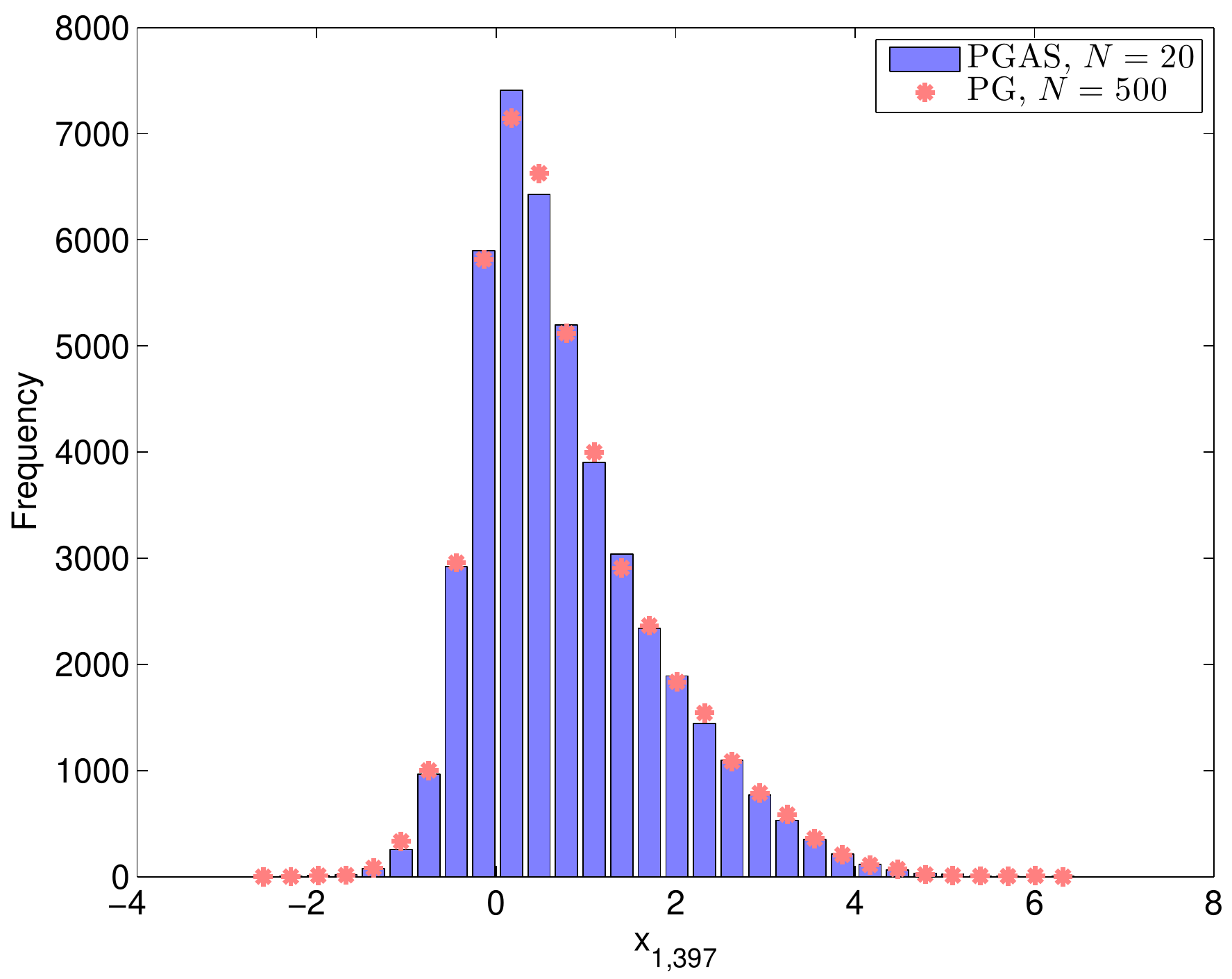}
  \caption{Posterior histograms for \pgas with $\Np = 20$ and $\ell = 4$ rejuvenated states (blue bars) and for \pg with $\Np = 500$ (pink asterisks) for a randomly chosen state, $x_{1,397}$.}
  \label{fig:degen-hist}
\end{figure}

We run the \pgas sampler with $\ell = 4$ rejuvenated states and with $\Np = 20$ particles.
To check that the sampler indeed converges to the correct posterior distribution we also run a \pg sampler
with $\Np = 500$ particles (this value was chosen by trial-and-error as the smallest number required by \pg
to still have reasonable mixing). In Figure~\ref{fig:degen-hist} we plot the histograms for the two samplers
for a randomly chosen state variable, $x_{1,397}$. As can be seen, there is a close match between the posterior
histograms.

\begin{figure}[ptb]
  \centering
  \includegraphics[width=0.7\linewidth]{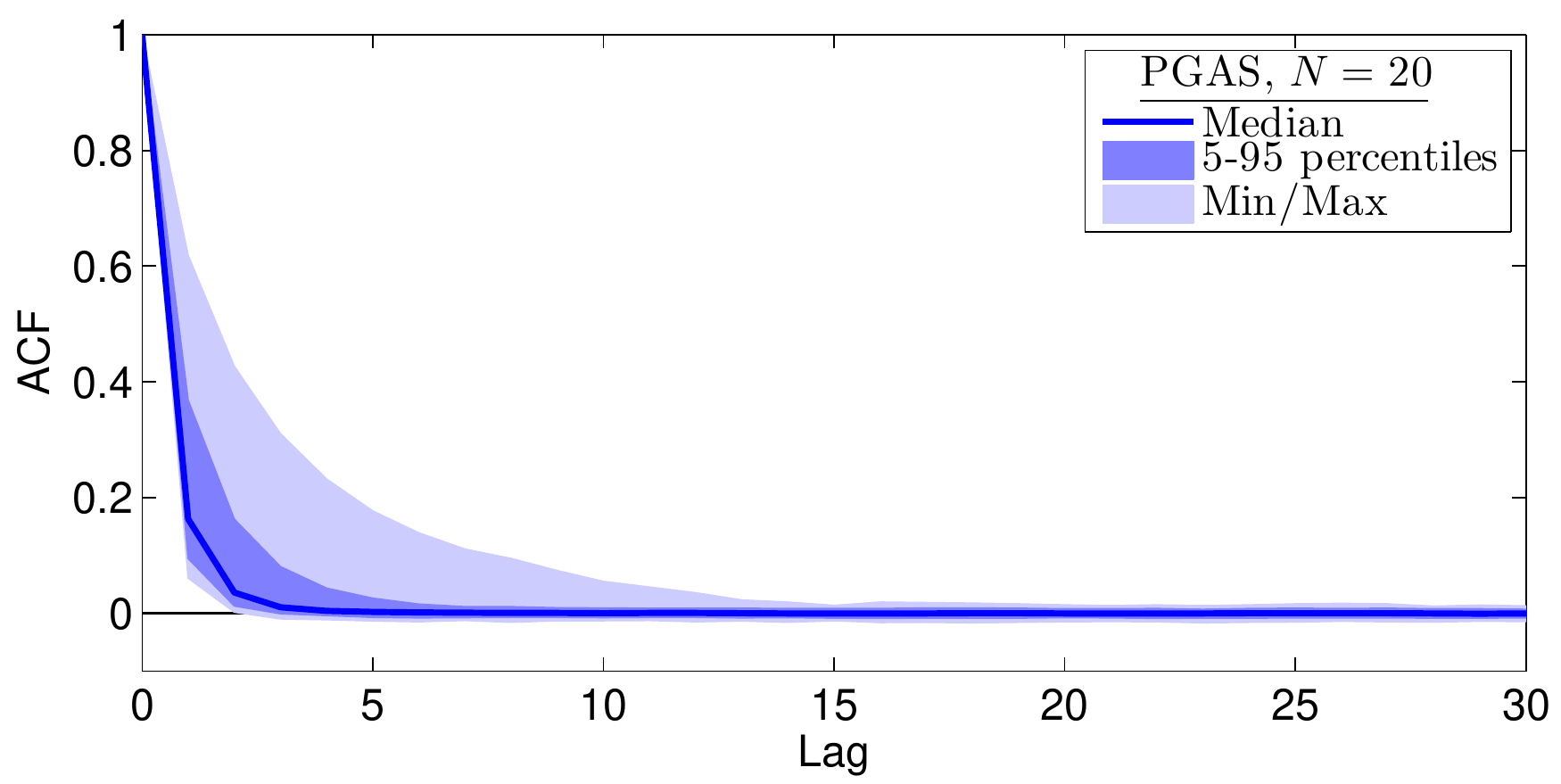}\\
  \includegraphics[width=0.7\linewidth]{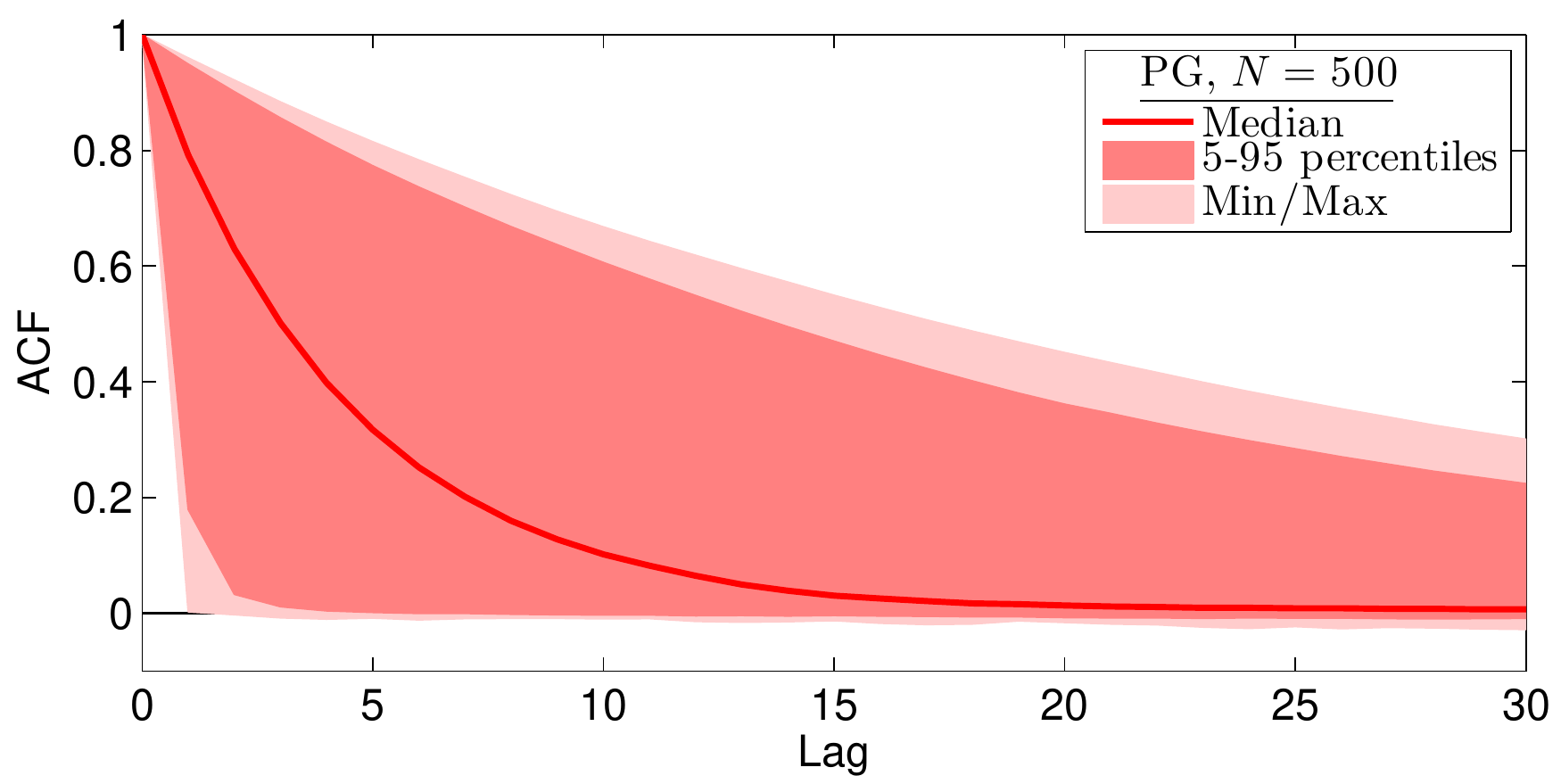}%
  \caption{Empirical autocorrelations for the state variables $\{x_{1,t}\}_{t=1}^{500}$ for \pgas with
    $\Np = 20$ and $\ell = 4$ rejuvenated states (top) and for \pg with $\Np = 500$ (bottom). The median (thick line),
  5-95 percentile (shaded area), and min/max-values (lighter shaded area) over the 500 state variables are reported.}
  \label{fig:degen-acf}
\end{figure}

We also compute the empirical autocorrelation functions for both samplers for all state variables $\{x_{1,t}\}_{t=1}^{500}$.
The results are reported in Figure~\ref{fig:degen-acf}. Despite the fact that it uses much fewer particles,
the mixing speed of \pgas is significantly better than for \pg. This is in agreement with previous results
reported in the literature \cite{LindstenJS:2014}. Indeed, the current example should mainly be seen
as an illustration of how particle rejuvenation opens up for using backward-sampling-based methods, in particular \pgas,
for a model where that would otherwise not be possible.

\subsection{Intractable transition model}\label{sec:numerical:abc}
We now turn to a model with an intractable transition density to
illustrate the \abc approximation for the \pgas sampler presented in Section~\ref{sec:abc}.
We consider inference in a
stochastic version of the Lorenz '63 model \cite{Lorenz:1963}, given by the following SDE:%
\begin{align}
  d
  \begin{bmatrix}
    Q_\ct \\ R_\ct \\ S_\ct
  \end{bmatrix}
  =
  \begin{bmatrix}
    \sigma (R_\ct - Q_\ct) \\ Q_\ct( \rho  - S_\ct ) - R_\ct \\ Q_\ct R_\ct - \beta S_\ct
  \end{bmatrix}dt
  +
  \begin{bmatrix}
    \sigma_Q & 0 & 0 \\
    0 & \sigma_R & 0 \\
    0 & 0 & \sigma_S
  \end{bmatrix}
  d W_\ct,
\end{align}
where $W_\ct$ is a three-dimensional Wiener process and the system parameters
are $\sigma = 10$, $\rho = 28$, $\beta = 8/3$ and $\sigma_Q=\sigma_R=\sigma_S = \sqrt{5}$.
The state is observed indirectly
through noisy observations of the $Q$-component at regular time intervals:
$y_t \sim \N(q_{t\Delta\ct}, 1)$ with $\Delta\ct = 0.01$. The initial state is distributed according
to $(Q_0, R_0, S_0)^\+ \sim \N(0,\eye{3})$.

A system simulator is implemented based on a fine-grid Milstein discretisation \cite{Milstein:1978}.
While the Milstein density for a single discretisation step is available \cite{Elerian:1998},
it is intractable to integrate out the intermediate steps on the grid.
Consequently, the employed simulator lacks a closed form transition density function and, indeed,
for the purpose of this illustration it is viewed simply as a ``black-box'' simulator.

We simulate the system for $\tau \in [0,10]$ and thus generate $\T = \thsnd{1}$ observations
$\prange{y_1}{y_\T}$. We then run \pgas with particle rejuvenation and the \abc approach outlined in Section~\ref{sec:abc}
to compute the posterior distribution of the system state at the observation time points.
The method uses $\Np=100$ particles and a Gaussian kernel for the \abc approximation:
\begin{align*}
  \ABCKernel{\epsilon}(x,x') = \exp\left(-\frac{\|x-x'\|^2}{2\epsilon}\right).
\end{align*}
 We let the
kernel bandwidth range from $\epsilon = 0.01$ to $\epsilon = 10$. As comparison, we also run both the \pg and \pimh
samplers from~\cite{AndrieuDH:2010} with the number of particles
$\Np$ ranging from 200 to \thsnd{10} (the computational cost per iteration
is roughly the same for \pgas with $\Np=100$ as for \pg/\pimh with $\Np = 200$, as the main computational cost comes
from the system simulator).

\begin{figure*}[ptb]
  \centering
  \includegraphics[width=0.5\linewidth]{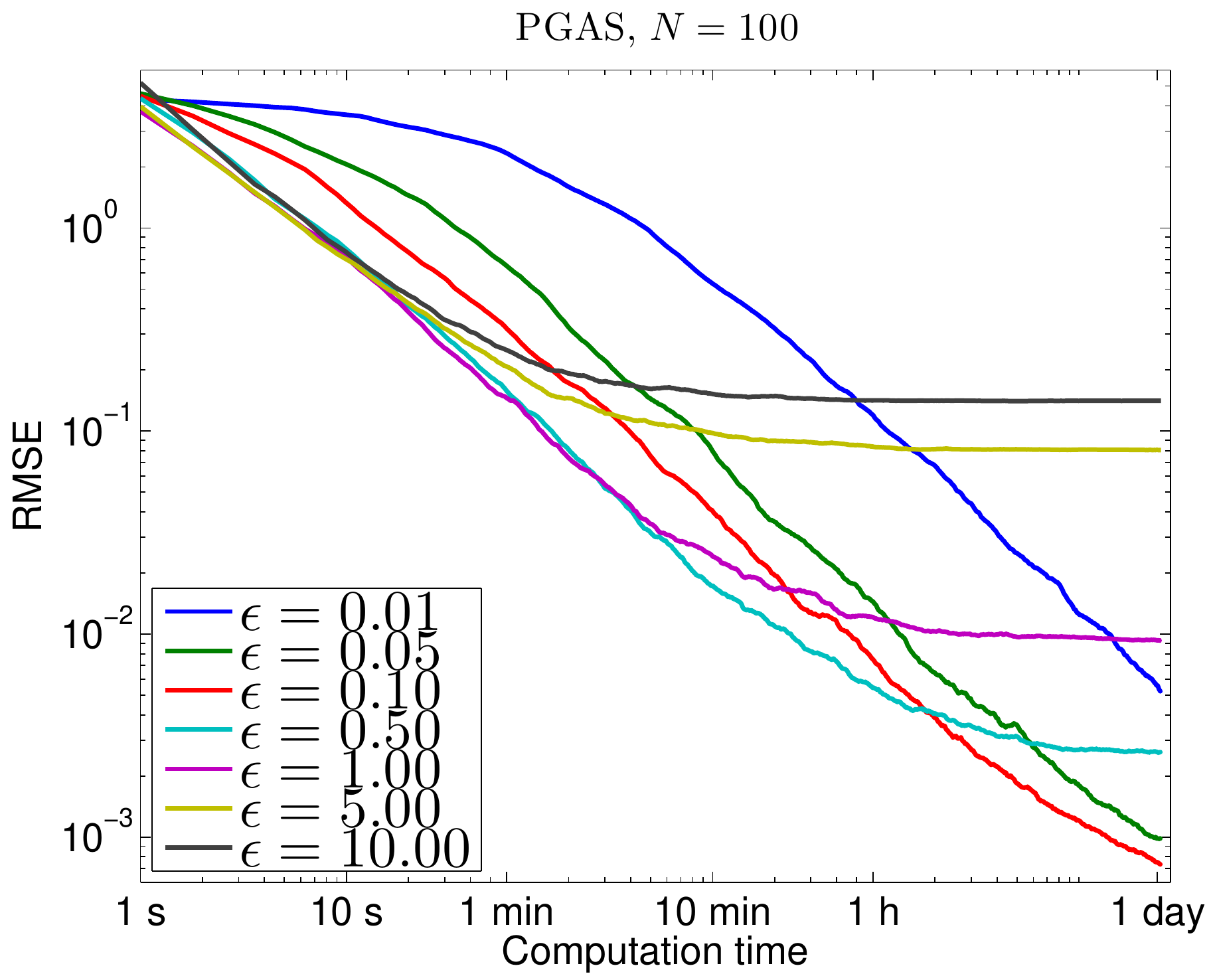}\\
  \includegraphics[width=0.5\linewidth]{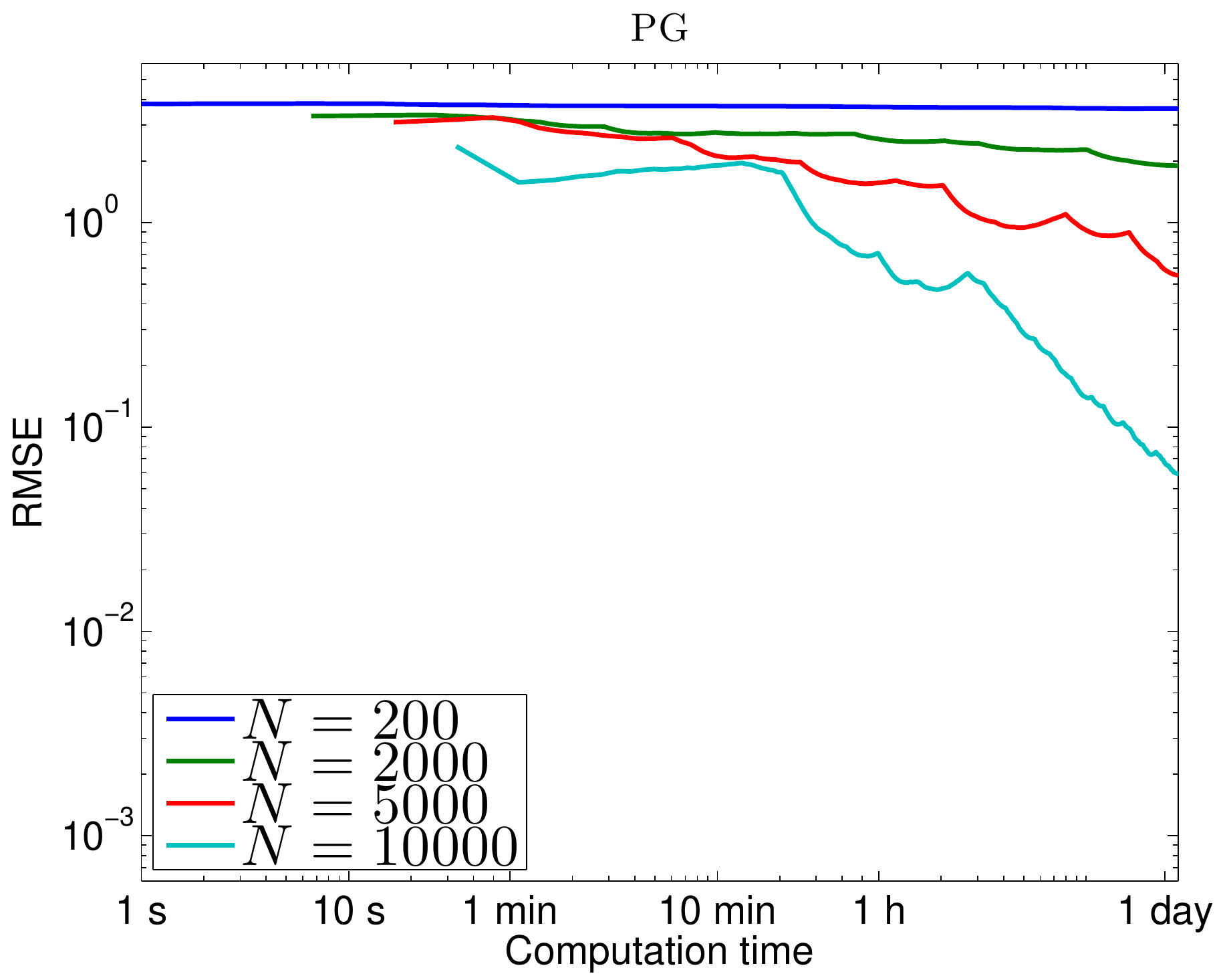}%
  \includegraphics[width=0.5\linewidth]{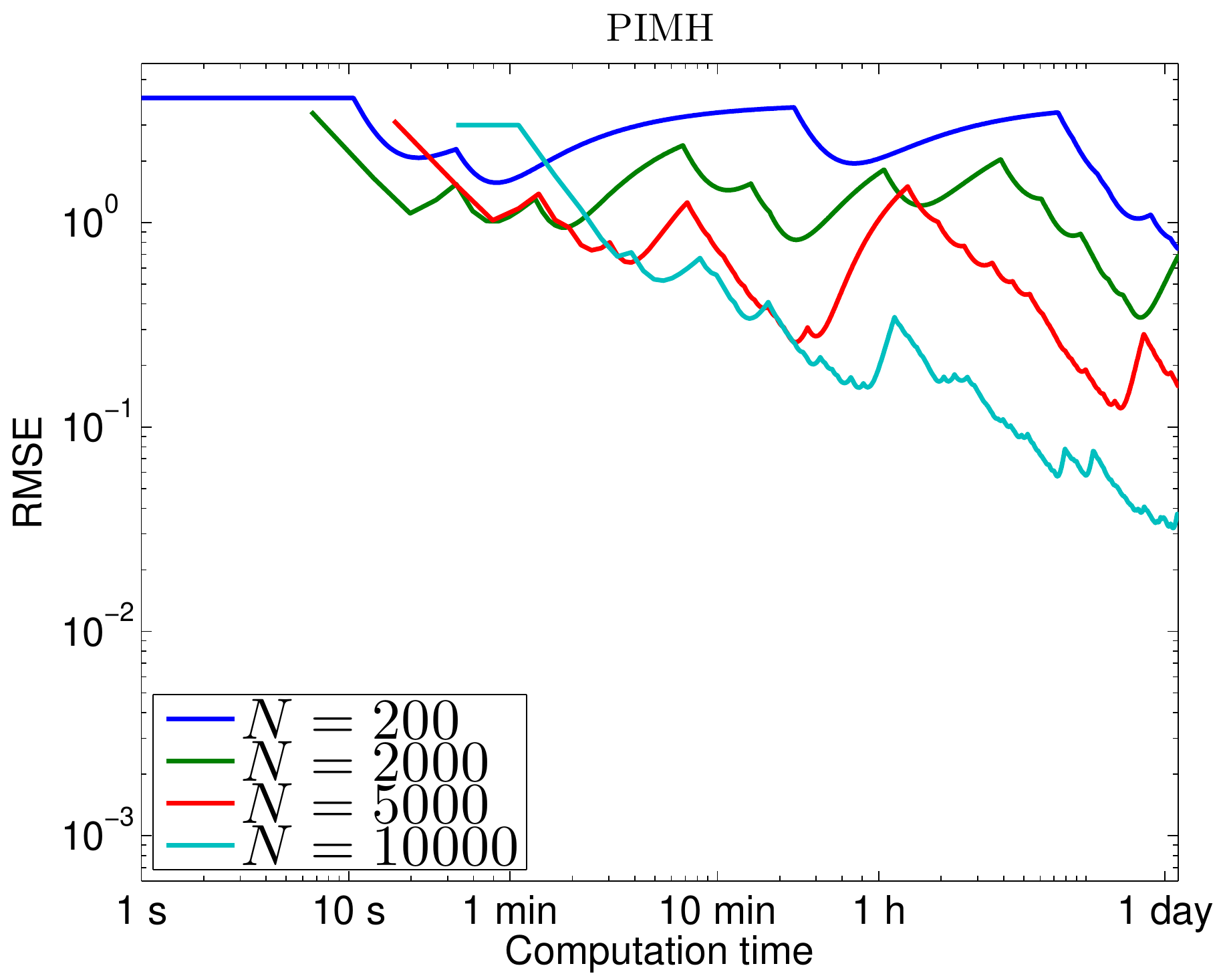}%
  \caption{RMSEs in the estimated posterior mean $\E[x_{1:\T} \mid y_{1:\T}]$ for the Lorenz '63 model for PGAS using $\Np = 100$ (top), \pg (bottom left), and \pimh (bottom right). (This figure is best viewed in color.)}
  \label{fig:lor63}
\end{figure*}

RMSEs for the posterior means of the system states are shown in Figure~\ref{fig:lor63}.%
\footnote{The ``ground truth'' is computed as an importance sampling estimator based on \thsnd{10} independent particle
filters, each one using $\Np = \thsnd{50}$ particles.}
The bias coming from the ABC approximation is evident for large $\epsilon$, as the RMSEs level out at a non-zero value (if
no approximation were made we would expect that the RMSE goes to zero\footnote{At least up to the accuracy of the ``ground
  truth'' reference sampler.} as the number of MCMC iterations increases).
Nevertheless, comparing the results for \pgas to those obtained for \pg and \pimh, it is evident that the ABC bias is
significantly smaller than the Monte Carlo errors resulting from the poor mixing of \pg and \pimh
(at least if $\epsilon$ is not overly large).

As $\epsilon$ decreases the bias diminishes,
but at the expense of slower convergence of PGAS. The reason for this is that the probability of updating the
ancestry decreases with $\epsilon$. In fact, for $\epsilon=0$ the bias is completely removed, but 
we will then have zero probability of changing the ancestry and \pgas will
be equivalent to \pg (and thus suffer from the same poor convergence speed).
Comparing the results for \pgas using $\epsilon = 0.01$ with \pg, however, we see that just having a small chance of
updating the ancestry can have a significant impact on the mixing speed. For such a small value of $\epsilon$
the \abc bias is clearly dominated by the variance, even after 24 hours of simulation, corresponding to roughly \thsnd{100} MCMC iterations.

\section{Discussion}
The particle rejuvenation technique presented in this paper generalises existing backward-simulation-based
methods and opens up for a high degree of flexibility when implementing these procedures.
This flexibility has been shown to be crucial for obtaining efficient samplers for several challenging types of
state space models with (nearly) degenerate and/or intractable transitions. However, the technique is more generally applicable
and we believe that it can be useful also for other types of models. In fact, we have recently made use of the particle rejuvenation
technique in a completely different setting, 
namely to prove the validity of the nested \smc algorithm presented in \cite{NaessethLS:2015}. To further investigate the
scope and usefulness of the particle rejuvenation technique in other contexts is a topic for future work.

Our main focus in this paper has been on state smoothing (or, more generally, inference for a latent stochastic process).
However, one of the main strengths of \pmcmc samplers, such as the \pgas algorithm that we have used
as the basis for the presented technique, is that they can be used for \emph{joint state and parameter} inference.
For \pgas, this is typically done by implementing a two-stage Gibbs sampler by iterating:
\begin{enumerate}
\item\label{step:1} Simulate the parameter $\theta$ from its full conditional given the states $\X_\T$ and observations $\Y_\T$.
\item Simulate the states $\X_\T$ from the \pgas Markov kernel, conditionally on $\theta$ and $\Y_\T$.
\end{enumerate}
This approach can be used also with the proposed Algorithm~\ref{alg:pgras} to obtain a valid MCMC sampler for
the joint posterior $p(\theta, \X_\T \mid \Y_\T)$. Indeed, this is the method that we used to sample the
parameter $\theta$ in the illustrative example in Section~\ref{sec:pgras:motivating}. However,
it is worth pointing out that this approach might not always be successful for the challenging model classes
\ma~and~\mb that have largely motivated the present development. The problem is that for these models it \emph{could}
be infeasible to simulate $\theta$ from its full conditional in Step~\ref{step:1} of the aforementioned Gibbs sampler,
since the degeneracy or intractability of the transition density \emph{could} be inherited by full conditional distribution of $\theta$.
In such scenarios, we thus need a different way for enabling the use of Algorithm~\ref{alg:pgras}
for parameter inference. We mention here two possible, albeit as of yet untested, approaches.

Firstly, even if the model is degenerate or intractable, it is typically possible to
explicitly introduce the ``noise variables'' $\{\drv_{t}\}_{t=1}^\T$ that drives the state
transition; see \eqref{eq:abc:degen_ssm}. We can then design a Gibbs (or Metropolis-within-Gibbs) sampler
for the extended model, with the original state variables $\X_{\T}$ marginalised out.
Note that we can still use Algorithm~\ref{alg:pgras} to simulate $\X_{\T}$, but
then transform the states to $\{ \drv_{t} \}_{t=1}^\T$ when updating $\theta$. This overcomes
the prohibitive $\Ordo(\T^2)$ computational complexity associated
with using \pgas for simulating the driving noise variables \emph{directly}, as discussed in~\cite{LindstenJS:2014}.

Secondly, it is possible to couple Algorithm~\ref{alg:pgras} with the particle marginal Metropolis-Hastings (\pmmh) algorithm
by \citet{AndrieuDH:2010}. \pmmh simulates $(\theta, \X_\T)$ jointly and implements a Metropolis-Hastings accept/reject
step based on an estimate of the data likelihood computed by running a (forward-in-time only) \smc sampler.
A problem with \pmmh, however, is that the method tends to get ``stuck'', due to occasional overestimation of the likelihood.
We believe that the method proposed in this paper can be used to mitigate this issue.
Indeed, it is possible to use Algorithm~\ref{alg:pgras} to refresh the likelihood estimate used in \pmmh,
while still maintaining the correct limiting distribution of the sampler (the details are omitted for brevity).
By occasionally refreshing the likelihood in this way, it may thus be possible to escape the sticky states
with overestimated likelihoods that deteriorate the practical performance of \pmmh.
%
%

Investigating the effectiveness of these approaches, as well as enabling parameter inference
in models of types \ma and \mb by using the method presented in Algorithm~\ref{alg:pgras} in a more direct sense, are topics for future work.

Another interesting and important direction for future work is to analyse the effect of the \abc approximation \eqref{eq:abc:kernel-approx}.
%
%
It was found empirically in \cite{LindstenJS:2014} that \pgas appears to be robust to approximation errors
in the ancestor sampling weights, and this is in agreements with our findings reported in Section~\ref{sec:numerical:abc}.
However, a more theoretical analysis is called for to understand if the sampler affected by the \abc approximation
still admits a limiting distribution and, if so, how this distribution is affected by the approximation error.

\appendix
\section{Partially collapsed Gibbs sampler}
The original \pgas method, reviewed in Algorithm~\ref{alg:pgas-original}, corresponds to the following
partially collapsed Gibbs sampler for the extended target distribution \eqref{eq:pgras:extended-target-def}; see \cite{LindstenJS:2014}:
Given $\X_{\T}  = \FX_1 = \X_\T'\in \setX^\T$ and $\B_{\T} = \FB_1 = \prange{\Np}{\Np}\in \crange{1}{\Np}^\T$:
\begin{enumerate}
\item[\gibbsnum{i}] Draw $\xx_{1}^{-b_{1}} \sim \extpdf( \,\cdot \mid \FX_1, \FB_1)$,
\item[\gibbsnum{ii}] For $t = 2$ to $\T$, draw:
  \begin{enumerate}
  \item[(a)] $ (\xx_t^{-b_t}, \aa_t^{ -b_t}) \sim \extpdf( \,\cdot \mid \XX_{t-1}, \AA_{t-1}, \FX_{t}, \FB_{t-1})$,
  \item[(b)] $ a_{t}^{b_{t}} \sim \extpdf( \,\cdot \mid \XX_{t-1}, \AA_{t-1}, \FX_{t}, \FB_{t})$,
  \end{enumerate}
\item[\gibbsnum{iii}] Draw $k \sim {}\extpdf( \,\cdot \mid \XX_{\T}, \AA_{\T})$.
\end{enumerate}
Similarly, the proposed \pgas algorithm with particle rejuvenation presented in Algorithm~\ref{alg:pgras} corresponds to
the following partially collapsed Gibbs sampler for \eqref{eq:pgras:extended-target-def}:
\begin{enumerate}
\item[\gibbsnum{i}]   Draw $\UX_{1} \sim \extpdf( \,\cdot \mid \RX_1, \FB_1)$,
\item[\gibbsnum{ii}]  Draw $\xx_{1}^{-b_{1}} \sim \extpdf( \,\cdot \mid \FX_1, \FB_1)$,
\item[\gibbsnum{iii}] For $t = 2$ to $\T$, draw:
  \begin{enumerate}
  \item[(a)] $(\xx_t^{-b_t}, \aa_t^{ -b_t}) \sim  \extpdf( \,\cdot \mid \XX_{t-1}, \AA_{t-1}, \FX_{t}, \FB_{t-1})$,
  \item[(b)] $(a_{t}^{b_{t}}, \UX_t) \sim \extpdf( \,\cdot \mid \XX_{t-1}, \AA_{t-1}, \RX_{t}, \FB_{t})$,
  \end{enumerate}
\item[\gibbsnum{iv}] Draw $k \sim {}\extpdf( \,\cdot \mid \XX_{\T}, \AA_{\T})$.
\end{enumerate}
More precisely, $\UX_1$ and $(a_{t}^{b_{t}}, \UX_t)$ are sampled from the Markov kernels $K_1$ and $K_t$, respectively, in
Steps $\gibbsnum{i}$ and $\gibbsnum{iii-b}$. However, since these Markov kernels are constructed to leave the
corresponding conditional distributions invariant, this corresponds to a standard composition of MCMC kernels.
The fact that the Gibbs sampler outlined above is properly collapsed, and thus leaves $\extpdf$ invariant,
follows by analogous arguments as in the proof of \cite[Theorem~1]{LindstenJS:2014}.

\bibliographystyle{plainnat}
\small
\bibliography{references}

\begin{thebibliography}{48}
\providecommand{\natexlab}[1]{#1}
\providecommand{\url}[1]{\texttt{#1}}
\expandafter\ifx\csname urlstyle\endcsname\relax
  \providecommand{\doi}[1]{doi: #1}\else
  \providecommand{\doi}{doi: \begingroup \urlstyle{rm}\Url}\fi

\bibitem[Andrieu and Vihola(2012)]{AndrieuV:2012}
C.~Andrieu and M.~Vihola.
\newblock Convergence properties of pseudo-marginal {M}arkov chain {M}onte
  {C}arlo algorithms.
\newblock \emph{The Annals of Applied Probability (forthcoming)}, 2012.
\newblock arXiv:1210.1484.

\bibitem[Andrieu et~al.(2003)Andrieu, de~Freitas, Doucet, and
  Jordan]{AndrieuFDJ:2003}
C.~Andrieu, N.~de~Freitas, A.~Doucet, and M.~I. Jordan.
\newblock An introduction to {MCMC} for machine learning.
\newblock \emph{Machine Learning}, 50\penalty0 (1):\penalty0 5--43, 2003.

\bibitem[Andrieu et~al.(2010)Andrieu, Doucet, and Holenstein]{AndrieuDH:2010}
C.~Andrieu, A.~Doucet, and R.~Holenstein.
\newblock Particle {M}arkov chain {M}onte {C}arlo methods.
\newblock \emph{Journal of the Royal Statistical Society: Series B},
  72\penalty0 (3):\penalty0 269--342, 2010.

\bibitem[Andrieu et~al.(2013)Andrieu, Lee, and Vihola]{AndrieuLV:2013}
C.~Andrieu, A.~Lee, and M.~Vihola.
\newblock Uniform ergodicity of the iterated conditional {SMC} and geometric
  ergodicity of particle {G}ibbs samplers.
\newblock arXiv.org, arXiv:1312.6432, December 2013.

\bibitem[Beaumont et~al.(2002)Beaumont, Zhang, and Balding]{BeaumontZB:2002}
M.~A. Beaumont, W.~Zhang, and D.~J. Balding.
\newblock Approximate {B}ayesian computation in population genetics.
\newblock \emph{Genetics}, 162\penalty0 (4):\penalty0 2025--2035, 2002.

\bibitem[Bunch and Godsill(2013)]{BunchG:2013}
P.~Bunch and S.~Godsill.
\newblock Improved particle approximations to the joint smoothing distribution
  using {M}arkov chain {M}onte {C}arlo.
\newblock \emph{{IEEE} Transactions on Signal Processing}, 61\penalty0
  (4):\penalty0 956--963, 2013.

\bibitem[Bunch et~al.(2015)Bunch, Lindsten, and Singh]{BunchLS:2015}
P.~Bunch, F.~Lindsten, and S.~S. Singh.
\newblock Particle {G}ibbs with refreshed backward simulation.
\newblock In \emph{Proceedings of the 40th {IEEE} International Conference on
  Acoustics, Speech and Signal Processing ({ICASSP})}, Brisbane, Australia,
  2015.
\newblock (accepted for publication).

\bibitem[Carter and Kohn(1994)]{CarterK:1994}
C.~K. Carter and R.~Kohn.
\newblock On {G}ibbs sampling for state space models.
\newblock \emph{Biometrika}, 81\penalty0 (3):\penalty0 541--553, 1994.

\bibitem[Carter et~al.(2014)Carter, Mendes, and Kohn]{CarterMK:2014}
C.~K. Carter, E.~F. Mendes, and R.~Kohn.
\newblock An extended space approach for particle {M}arkov chain {M}onte
  {C}arlo methods.
\newblock arXiv.org, arXiv:1406.5795, July 2014.

\bibitem[Chopin and Singh(2014)]{ChopinS:2014}
N.~Chopin and S.~S. Singh.
\newblock On particle {G}ibbs sampling.
\newblock \emph{Bernoulli}, 2014.
\newblock Forthcoming.

\bibitem[Dean et~al.(2014)Dean, Singh, Jasra, and Peters]{DeanSJP:2014}
T.~A. Dean, S.~S. Singh, A.~Jasra, and G.~W. Peters.
\newblock Parameter estimation for hidden {M}arkov models with intractable
  likelihoods.
\newblock \emph{The Scandinavian Journal of Statistics}, 41\penalty0
  (4):\penalty0 970--987, 2014.

\bibitem[Del~Moral(2004)]{DelMoral:2004}
P.~Del~Moral.
\newblock \emph{{F}eynman-{K}ac Formulae - Genealogical and Interacting
  Particle Systems with Applications}.
\newblock Probability and its Applications. Springer, 2004.

\bibitem[Del~Moral et~al.(2006)Del~Moral, Doucet, and Jasra]{DelMoralDJ:2006}
P.~Del~Moral, A.~Doucet, and A.~Jasra.
\newblock Sequential {M}onte {C}arlo samplers.
\newblock \emph{Journal of the Royal Statistical Society: Series {B}},
  68\penalty0 (3):\penalty0 411--436, 2006.

\bibitem[Del~Moral et~al.(2014)Del~Moral, Kohn, and Patras]{DelMoralKP:2014}
P.~Del~Moral, R.~Kohn, and F.~Patras.
\newblock On particle {G}ibbs {M}arkov chain {M}onte {C}arlo models.
\newblock arXiv.org, arXiv:1404.5733, 2014.

\bibitem[Dempster et~al.(1977)Dempster, Laird, and Rubin]{DempsterLR:1977}
A.~Dempster, N.~Laird, and D.~Rubin.
\newblock Maximum likelihood from incomplete data via the {EM} algorithm.
\newblock \emph{Journal of the {R}oyal {S}tatistical {S}ociety, {S}eries {B}},
  39\penalty0 (1):\penalty0 1--38, 1977.

\bibitem[Doucet and Johansen(2011)]{DoucetJ:2011}
A.~Doucet and A.~Johansen.
\newblock A tutorial on particle filtering and smoothing: Fifteen years later.
\newblock In D.~Crisan and B.~Rozovskii, editors, \emph{The Oxford Handbook of
  Nonlinear Filtering}, pages 656--704. Oxford University Press, Oxford, UK,
  2011.

\bibitem[Doucet et~al.(2014)Doucet, Pitt, Deligiannidis, and
  Kohn]{DoucetPDK:2014}
A.~Doucet, M.~K. Pitt, G.~Deligiannidis, and R.~Kohn.
\newblock Efficient implementation of {M}arkov chain {M}onte {C}arlo when using
  an unbiased likelihood estimator.
\newblock \emph{Biometrika (forthcoming)}, 2014.
\newblock Preprint, arXiv:1210.1871v3.

\bibitem[Durham and Gallant(2002)]{DurhamG:2002}
G.~B. Durham and A.~R. Gallant.
\newblock Numerical techniques for maximum likelihood estimation of
  continuous-time diffusion processes.
\newblock \emph{Journal of Business \& Economic Statistics}, 20\penalty0
  (3):\penalty0 297--316, 2002.

\bibitem[Dyk and Park(2008)]{DykP:2008}
D.~A.~Van Dyk and T.~Park.
\newblock Partially collapsed {G}ibbs samplers: Theory and methods.
\newblock \emph{Journal of the American Statistical Association}, 103\penalty0
  (482):\penalty0 790--796, 2008.

\bibitem[Elerian(1998)]{Elerian:1998}
O.~Elerian.
\newblock A note on the existence of a closed form conditional transition
  density for the {M}ilstein scheme.
\newblock Economics Discussion Paper 1998-W18, Nuffield College, Oxford, 1998.

\bibitem[Fearnhead and Prangle(2012)]{FearnheadP:2012}
P.~Fearnhead and D.~Prangle.
\newblock Constructing summary statistics for approximate {B}ayesian
  computation: semi-automatic approximate {B}ayesian computation.
\newblock \emph{Journal of the Royal Statistical Society: Series {B}},
  74\penalty0 (3):\penalty0 419--474, 2012.

\bibitem[Fr\"uhwirth-Schnatter(1994)]{Fruhwirth-Schnatter:1994}
S.~Fr\"uhwirth-Schnatter.
\newblock Data augmentation and dynamic linear models.
\newblock \emph{Journal of Time Series Analysis}, 15\penalty0 (2):\penalty0
  183--202, 1994.

\bibitem[Godsill and Rayner(1998)]{Godsill1998}
S.~Godsill and P.~Rayner.
\newblock \emph{Digital audio restoration}.
\newblock Springer, 1998.

\bibitem[Godsill et~al.(2004)Godsill, Doucet, and West]{GodsillDW:2004}
S.~J. Godsill, A.~Doucet, and M.~West.
\newblock {M}onte {C}arlo smoothing for nonlinear time series.
\newblock \emph{Journal of the American Statistical Association}, 99\penalty0
  (465):\penalty0 156--168, March 2004.

\bibitem[Golightly and Wilkinson(2011)]{GolightlyW:2011}
A.~Golightly and D.~J. Wilkinson.
\newblock Bayesian parameter inference for stochastic biochemical network
  models using particle {M}arkov chain {M}onte {C}arlo.
\newblock \emph{Interface Focus}, 1\penalty0 (6):\penalty0 807--820, 2011.

\bibitem[Gustafsson et~al.(2002)Gustafsson, Gunnarsson, Bergman, Forssell,
  Jansson, Karlsson, and Nordlund]{GustafssonGBFJKN:2002}
F.~Gustafsson, F.~Gunnarsson, N.~Bergman, U.~Forssell, J.~Jansson, R.~Karlsson,
  and P.-J. Nordlund.
\newblock Particle filters for positioning, navigation, and tracking.
\newblock \emph{{IEEE} Transactions on Signal Processing}, 50\penalty0
  (2):\penalty0 425--437, 2002.

\bibitem[Handschin and Mayne(1969)]{HandschinM:1969}
J.~Handschin and D.~Mayne.
\newblock {M}onte {C}arlo techniques to estimate the conditional expectation in
  multi-stage non-linear filtering.
\newblock \emph{International Journal of Control}, 9\penalty0 (5):\penalty0
  547--559, May 1969.

\bibitem[Kailath et~al.(2000)Kailath, Sayed, and Hassibi]{KailathSH:2000}
T.~Kailath, A.~H. Sayed, and B.~Hassibi.
\newblock \emph{Linear Estimation}.
\newblock Prentice Hall, Upper Saddle River, NJ, USA, 2000.

\bibitem[Li and Jilkov(2003)]{Li2003}
X~R Li and V~P Jilkov.
\newblock Survey of maneuvering target tracking. part {I}: Dynamic models.
\newblock \emph{IEEE Transactions on Aerospace and Electronic Systems},
  39\penalty0 (4):\penalty0 1333--1364, 2003.

\bibitem[Lindsten and Sch\"on(2013)]{LindstenS:2013}
F.~Lindsten and T.~B. Sch\"on.
\newblock Backward simulation methods for {M}onte {C}arlo statistical
  inference.
\newblock \emph{Foundations and Trends in Machine Learning}, 6\penalty0
  (1):\penalty0 1--143, 2013.

\bibitem[Lindsten et~al.(2014)Lindsten, Jordan, and Sch\"on]{LindstenJS:2014}
F.~Lindsten, M.~I. Jordan, and T.~B. Sch\"on.
\newblock Particle {G}ibbs with ancestor sampling.
\newblock \emph{Journal of Machine Learning Research}, 15:\penalty0 2145--2184,
  2014.

\bibitem[Lindsten et~al.(2015)Lindsten, Douc, and Moulines]{LindstenDM:2015}
F.~Lindsten, R.~Douc, and E.~Moulines.
\newblock Uniform ergodicity of the particle {G}ibbs sampler.
\newblock \emph{Scandinavian Journal of Statistics (forthcoming)}, 2015.
\newblock \doi{10.1111/sjos.12136}.
\newblock Preprint, arXiv:1401.0683.

\bibitem[Lorenz(1963)]{Lorenz:1963}
E.~N. Lorenz.
\newblock Deterministic nonperiodic flow.
\newblock \emph{Journal of the Atmospheric Sciences}, 20\penalty0 (2):\penalty0
  130--141, 1963.

\bibitem[McKinley et~al.(2009)McKinley, Cook, and Deardon]{McKinleyCD:2009}
T.~McKinley, A.~R. Cook, and R.~Deardon.
\newblock Inference in epidemic models without likelihoods.
\newblock \emph{The International Journal of Biostatistics}, 5\penalty0
  (1):\penalty0 1557--4679, 2009.

\bibitem[Milstein(1978)]{Milstein:1978}
G.~N. Milstein.
\newblock A method of second-order accuracy integration of stochastic
  differential equations.
\newblock \emph{Theory of Probability and its Applications}, 23:\penalty0
  396--401, 1978.

\bibitem[Murray et~al.(2013)Murray, Jones, and Parslow]{MurrayJP:2013}
L.~M. Murray, E.~M. Jones, and J.~Parslow.
\newblock On disturbance state-space models and the particle marginal
  {M}etropolis-{H}astings sampler.
\newblock \emph{SIAM/ASA Journal on Uncertainty Quantification}, 1\penalty0
  (1):\penalty0 494--521, 2013.

\bibitem[Naesseth et~al.(2015)Naesseth, Lindsten, and Sch\"on]{NaessethLS:2015}
C.~A. Naesseth, F.~Lindsten, and T.~B. Sch\"on.
\newblock Nested sequential {M}onte {C}arlo methods.
\newblock arXiv.org, arXiv:1502.02536, February 2015.

\bibitem[Olsson and Ryd\'en(2010)]{OlssonR:2010}
J.~Olsson and T.~Ryd\'en.
\newblock Metropolising forward particle filtering backward sampling and
  {R}ao-{B}lackwellisation of {M}etropolised particle smoothers.
\newblock Technical Report 2010:15, Mathematical Sciences, Lund University,
  Lund, Sweden, 2010.

\bibitem[Pitt et~al.(2012)Pitt, Silva, Giordani, and Kohn]{PittSGK:2012}
M.~K. Pitt, R.~S. Silva, P.~Giordani, and R.~Kohn.
\newblock On some properties of {M}arkov chain {M}onte {C}arlo simulation
  methods based on the particle filter.
\newblock \emph{Journal of Econometrics}, 171:\penalty0 134--151, 2012.

\bibitem[Rasmussen et~al.(2011)Rasmussen, Ratmann, and
  Koelle]{RasmussenRK:2011}
D.~A. Rasmussen, O.~Ratmann, and K.~Koelle.
\newblock Inference for nonlinear epidemiological models using genealogies and
  time series.
\newblock \emph{PLoS Comput Biology}, 7\penalty0 (8), 2011.

\bibitem[Robert and Casella(2004)]{RobertC:2004}
C.~P. Robert and G.~Casella.
\newblock \emph{{M}onte {C}arlo Statistical Methods}.
\newblock Springer, 2004.

\bibitem[Rubin(1987)]{Rubin:1987}
D.~B. Rubin.
\newblock A noniterative sampling/importance resampling alternative to the data
  augmentation algorithm for creating a few imputations when fractions of
  missing information are modest: The {SIR} algorithm.
\newblock \emph{Journal of the American Statistical Association}, 82\penalty0
  (398):\penalty0 543--546, June 1987.
\newblock Comment to Tanner and Wong: The Calculation of Posterior
  Distributions by Data Augmentation.

\bibitem[Tanner and Wong(1987)]{TannerW:1987}
M.~A. Tanner and W.~H. Wong.
\newblock The calculation of posterior distributions by data augmentation.
\newblock \emph{Journal of the American Statistical Association}, 82\penalty0
  (398):\penalty0 528--540, June 1987.

\bibitem[Tavar{\'e} et~al.(1997)Tavar{\'e}, Balding, Griffiths, and
  Donnelly]{TavareBGD:1997}
S.~Tavar{\'e}, D.~J. Balding, R.~C. Griffiths, and P.~Donnelly.
\newblock Inferring coalescence times from {DNA} sequence data.
\newblock \emph{Genetics}, 145\penalty0 (2):\penalty0 505--518, 1997.

\bibitem[Tierney(1994)]{Tierney:1994}
L.~Tierney.
\newblock {M}arkov chains for exploring posterior distributions.
\newblock \emph{The Annals of Statistics}, 22\penalty0 (4):\penalty0
  1701--1728, 1994.

\bibitem[Vrugt et~al.(2013)Vrugt, ter Braak, Diks, and Schoups]{VrugtBDS:2013}
J.~A. Vrugt, J.~F. ter Braak, C.~G.~H. Diks, and G.~Schoups.
\newblock Hydrologic data assimilation using particle {M}arkov chain {M}onte
  {C}arlo simulation: {T}heory, concepts and applications.
\newblock \emph{Advances in Water Resources}, 51:\penalty0 457--478, 2013.

\bibitem[Whiteley(2010)]{Whiteley:2010}
N.~Whiteley.
\newblock Discussion on {P}article {M}arkov chain {M}onte {C}arlo methods.
\newblock \emph{Journal of the Royal Statistical Society: Series {B}},
  72\penalty0 (3):\penalty0 306--307, 2010.

\bibitem[Whiteley et~al.(2010)Whiteley, Andrieu, and Doucet]{WhiteleyAD:2010}
N.~Whiteley, C.~Andrieu, and A.~Doucet.
\newblock Efficient {B}ayesian inference for switching state-space models using
  discrete particle {M}arkov chain {M}onte {C}arlo methods.
\newblock Technical Report Bristol Statistics Research Report 10:04, University
  of Bristol, 2010.

\end{thebibliography}

\end{document}